%% file: test.tex
\documentclass[a4paper,UKenglish]{lipics-v2018}
\usepackage{microtype}%if unwanted, comment out or use option "draft"

\usepackage{xcolor}
\usepackage{soul}
\usepackage[utf8]{inputenc}
\usepackage{caption}
\usepackage{array}

\usepackage{amssymb}
\usepackage{helvet}
\usepackage{courier}
\usepackage{tcolorbox}
\usepackage{amsmath}
\usepackage{amsfonts}
\usepackage{verbatim}
\usepackage{bbm}
\usepackage{amsthm}

\usepackage{xspace}
\usepackage{url}
\usepackage{mathrsfs}
\usepackage{graphicx}
\usepackage{amsmath}
\usepackage{amssymb}
\usepackage{amsthm}
\usepackage{multirow} 
\usepackage{subfloat}
\usepackage{tikz}
\usetikzlibrary{arrows,decorations,decorations.shapes,decorations.markings,decorations.pathreplacing,backgrounds,shapes,petri,topaths}
\usepackage{tkz-berge}
\usepackage{pgfplots}
\usepackage{multirow}
\usepackage{enumitem}

\usepackage{graphics}

\newtheorem{conjecture}[theorem]{Conjecture}
\newtheorem{proposition}[theorem]{Proposition}
\newtheorem{observation}[theorem]{Observation}
\usepackage{todonotes}
\input{defs.tex}

\def\ol{\overline}

\newcount\Comments  % 0 suppresses notes to selves in text
\definecolor{darkgreen}{rgb}{0,0.6,0}
\newcommand{\kibitz}[2]{\ifnum\Comments=1{\color{#1}{#2}}\fi}
\newcommand{\rmr}[1]{\kibitz{blue}{[RESHEF:#1]}}
\newcommand{\adl}[1]{\kibitz{red}{[ARGY:#1]}}

\newcommand{\NP}{$\mathcal{NP}$\xspace}
\newcommand{\PP}{$\mathcal{P}$\xspace}
\newcommand{\dminor}{\textsc{IsDMinor}\xspace}

\newcommand{\tpath}{2-\textsc{Path}\xspace}

\newcommand{\maxdcut}{\textsc{MaxDiCut}\xspace}

\newcommand{\iss}{\textsc{IsSerial}\xspace}
\newcommand{\isc}{\textsc{IsConcurrent}\xspace}
\newcommand{\isp}{\textsc{IsParallel}\xspace}
\newcommand{\issc}{\textsc{IsSerialConcurrent}\xspace}
\newcommand{\issp}{\textsc{IsSerialParallel}\xspace}
\newcommand{\maxs}{\textsc{MaxSerial}\xspace}
\newcommand{\maxc}{\textsc{MaxConcurrent}\xspace}
\newcommand{\maxp}{\textsc{MaxParallel}\xspace}
\newcommand{\maxsc}{\textsc{MaxSerialConcurrent}\xspace}
\newcommand{\maxsp}{\textsc{MaxSerialParallel}\xspace}

\newcommand{\noijcai}[1]{}

\usepackage[numbers,sort]{natbib}

\title{Directed Graph Minors and Serial-Parallel Width}

\titlerunning{Directed Graph Minors and Serial-Parallel Width}%optional, please use if title is longer than one line

\author{Argyrios Deligkas}{Leverhulme Research Centre, University of Liverpool, UK}{argyrios.deligkas@liverpool.ac.uk}{}{}%{https://orcid.org/0000-0002-1825-0097}{[funding]}
\author{Reshef Meir}{Faculty of Industrial Engineering and Management, Technion, Israel.}{reshefm@technion.ac.il}{}{}%{[orcid]}{[funding]}

\authorrunning{A.\, Deligkas and R.\, Meir}%mandatory. First: Use abbreviated first/middle names. Second (only in severe cases): Use first author plus 'et al.'

\Copyright{Argyrios Deligkas and Reshef Meir}%mandatory, please use full first names. LIPIcs license is "CC-BY";  http://creativecommons.org/licenses/by/3.0/

% mandatory: Please choose ACM 2012 classifications from https://www.acm.org/publications/class-2012 or https://dl.acm.org/ccs/ccs_flat.cfm . E.g., cite as "General and reference $\rightarrow$ General literature" or \ccsdesc[100]{General and reference~General literature}. 

%mandatory

\category{}%optional, e.g. invited paper

\relatedversion{}%optional, e.g. full version hosted on arXiv, HAL, or other respository/website

\supplement{}%optional, e.g. related research data, source code, ... hosted on a repository like zenodo, figshare, GitHub, ...

\funding{}%optional, to capture a funding statement, which applies to all authors. Please enter author specific funding statements as fifth argument of the \author macro.

\EventEditors{Igor Potapov, Paul Spirakis, and James Worrell}
\EventNoEds{3}
\EventLongTitle{43rd (MFCS 2018)}
\EventShortTitle{MFCS 2018}
\EventAcronym{MFCS}
\EventYear{2018}
\EventDate{August 27--31, 2018}
\EventLocation{Liverpool, United Kingdom}
\EventLogo{}
\SeriesVolume{43}
\ArticleNo{1}
\nolinenumbers %uncomment to disable line numbering
\hideLIPIcs  %uncomment to remove references to LIPIcs series (logo, DOI, ...), e.g. when preparing a pre-final version to be uploaded to arXiv or another public repository
\usepackage{fancyhdr}
\begin{document}

\maketitle

\begin{abstract}
Graph minors are a primary tool in understanding the structure of undirected graphs, with many conceptual and algorithmic implications.
We propose new variants of \emph{directed graph minors} and \emph{directed graph embeddings}, by modifying familiar definitions.
For the class of 2-terminal directed acyclic graphs (TDAGs) our two definitions coincide, and the class is closed under both operations. The usefulness of our directed minor operations
is demonstrated by characterizing all TDAGs with serial-parallel width at most $k$;
a class of networks known to guarantee bounded negative externality in nonatomic routing games. Our characterization implies that a TDAG has serial-parallel width of $1$ if and only if it is a directed series-parallel graph.
We also study the computational complexity of finding a directed minor and computing the serial-parallel width.       
 \end{abstract}

\section{Introduction}
Graph theory has been one of the fundamental tools in artificial intelligence since its inception and in many AI challenges the inputs are in a form of a graph, e.g., analysis of electric circuits and communication networks, and training of neural nets. %~\cite{codenotti1991parallel}, routing in wired and wireless networks~\cite{ahlswede2000network,carre1971algebra,kulkarni2011computational}, training a neural network~\cite{frean1990upstart}. 
More important still, numerous problems from all domains in AI are often solved by reducing them to some algorithmic problem on a graph. Some prominent examples include search and path-finding~\cite{pohl1970heuristic}; planning graphs~\cite{blum1997fast}; constraint satisfaction~\cite{mackworth1981consistency}; AND-OR graph~\cite{chang1971admissible}; %and SUM-PRODUCT graphs~\cite{chang1971admissible,poon2011sum}; 
and inference in Bayesian networks~\cite{cooper1990computational}.

The \emph{structure} of these graphs is often crucial to the modeling of the problem. For instance, the last two examples above use \emph{directed acyclic graphs} (DAGs), which are also used to represent belief structures, influence relations and decision diagrams~\cite{horvitz1988decision}. Restrictions on the degree, maximum length, or other properties of the underlying graph, can be exploited: problems that are not guaranteed to have a solution in general may behave better on some classes of graphs, and many algorithms are guaranteed to have a lower runtime subject to structural assumptions. 
%\vspace{-3mm}
\subsection*{Graph minors}
When considering \emph{undirected graphs}, some of the primary tools of structural analysis use \emph{graph embeddings} and \emph{graph minors}. These are substructures whose exclusion from a graph indicates certain ``simplicity'' properties. % have been used extensively in order to characterize classes of graphs with various properties.
 Some famous results are the characterization of planar graphs~\cite{kuratowski1930probleme}, and of graphs with bounded treewidth~\cite{robertson1986graph} by excluded minors. In fact, for undirected graphs there is by now a sound theory of graph minors with many applications; see, e.g., \cite{lovasz2006graph} for a  survey, and  \cite{demaine2005algorithmic} for algorithmic implications. The culmination of this theory is the Graph Minor Theorem~\cite{wagner1937eigenschaft,robertson2004graph}, which states that \emph{any} class of undirected graphs that is closed under the minor operation, can be characterized by a finite set of excluded minors.  

%Perhaps the most important application to artificial intelligence is the use of graph minor theory to develop efficient algorithms 
Perhaps the most important application of graph minor theory to AI  is its use for developing efficient algorithms 
on graphs with bounded treewidth and/or other properties~\cite{gogate2004complete,cohen2015tractable,rowland2017conditions}. Graph minors were also used to characterize classes of graphs induced by planning problems to identify potential effects of time-inconsistent planning~\cite{kleinberg2014time,tang2017computational}.

\medskip
Although the graphs encountered in many AI problems are \emph{directed}, there is  no single theory of directed graph minors, and results are far more scarce than in the undirected case. Several papers suggested various definitions of directed minors, embeddings, and subdivisions, and provided various characterization results~\cite{johnson2001directed,holzman2003network,kintali2013forbidden,holzman2003network,johnson2015excluding,kawarabayashi2015towards,kreutzer2016nowhere}. However, each such definition uses different graph operations, some of which we explain in detail later on. Certain notions of directed minors are only applicable for subclasses of directed graphs. For example, the definitions in \cite{kreutzer2016nowhere} apply only to minors with a certain structure called ``crown''. 

In this paper, we will be interested mainly in directed graphs that are acyclic (DAG), or 2-terminal, or both (TDAG). 2-terminal graphs occur in routing~\cite{ashlagi2009two}, circuit analysis~\cite{shannon1949synthesis} and in many planning problems~\cite{kleinberg2014time}. Thus  understanding the structure of graphs in these classes is an important challenge. 

\subparagraph*{\bf Paper structure and contribution.}
In the first part of the paper (Section~\ref{sec:minors}) we define new notions of graph embedding and graph minor for general directed graphs.\footnote{To avoid confusion, we should note that the term \emph{graph embedding} is  used in the machine learning literature to describe embedding of graphs in various topological or metric spaces (e.g., \cite{wang2014knowledge}), which is a very different problem.} We show that for the class of 2-terminal directed acyclic graphs (TDAGs) these two operations exactly reverse one another. Thus, a TDAG $G'$ is a directed minor of $G$ if and only if it is embedded in $G$.  Also, the class of TDAG is closed under directed minor and directed embedding operations. We thus argue that our definitions provide a sound basis for a theory of directed graph minors, at least for the class of TDAGs. 

To demonstrate the usefulness of our directed minor theory, we apply it in Section~\ref{sec:SPW} to characterize TDAGs with bounded \emph{parallel width} and \emph{serial-parallel width}. The parallel width of a graph corresponds to the maximal cut separating the source from the target. 
Serial-parallel width of a graph is a parameter recently introduced in the context of routing games~\cite{MP18}, and it is useful for bounding negative externalities.  For example, series-parallel graphs have serial-parallel width of 1, and the famous Braess paradox network has a serial-parallel width of 2, which intuitively means that there is a route intersecting two other  routes.
\adl{the explanation is not completely clear.}
 %thus a group of agents can increase the congestion for two other groups of agents.
We describe a finite set of graphs (generalized variants of the Braess/Wheatstone network) whose exclusion as directed minors of a TDAG $G$ is necessary and sufficient to determine that $G$ has serial-parallel width lower than $k$, for any $k$. 
For example, for $k=1$ there is only one forbidden  directed minor, which is the Braess paradox graph. Thus, our results extend known results for directed and undirected graphs~\cite{holzman2003network,milchtaich2006network}. %, albeit with somewhat different notions of graph embeddings. 

In Section~\ref{sec:comp} we settle several computational questions arising from our definitions. 
\noijcai{
For example, it is \NP-complete to decide whether a given graph $G'$ is a directed minor of another graph $G$, even when both graphs are TDAGs, but a polynomial algorithm exists when the It is also \NP-hard to compute the parallel width of a 2-terminal graph. However,  we design polynomial algorithms for all these problems when $G$ is a TDAG and $k$, or the size of $G'$, is fixed. Some  proofs are omitted and appear in the appendix.}

%%%%%%+++++++++++++++++
%%%%%%+++++++++++++++++

\section{Preliminaries}
\label{sec:prelim}
For convenience, we will use the letter $H$ for undirected graphs, and the letter $G$ for directed graphs. We denote by $\ol G$ the undirected graph obtained from $G$ by ignoring edge directions. 
We denote a path in graph $\tup{V,E}$ by $(v_1,v_2,\ldots,v_m)$, where for every $i\leq m-1$, $(v_i,v_{i+1})\in E$. We use dash to abbreviate the path, e.g. $a-b-c$ is an abbreviation to a path $(a,\ldots,b,\ldots,c)$.
 
If  nodes $x,y$ are on some  path $p$, then $p_{xy}$ denotes the \emph{open} subpath of $p$ between nodes $x$ and $y$, and $[p_{xy}]=x-p_{xy}-y$ the \emph{closed} subpath that includes the extreme vertices. 
Thus, for example if $p=(a_1,a_2,a_3,a_4,a_5,a_6)$ we can write $p=a_1-[p_{a_2 a_4}]-a_5-a_6$ or $p=a_1-p_{a_1 a_4} - a_4 - [p_{a_5 a_6}]$. We denote by $|p|$ the number of edges in $p$, thus $|[p_{ab}]|=|p_{ab}|$.

\begin{definition}[2-terminal graph~\cite{milchtaich2006network,holzman2003network}]
A  \emph{2-terminal [directed] graph} $G=\tup{V,E,s,t}$ is a [directed] multigraph with no self-loops and two distinguished vertices $s,t\in V$, such that every vertex and edge belong to at least one [directed] simple  $s-t$ path. 
\end{definition}

A \emph{forward-subtree} of a directed 2-terminal graph $G$ is a subset of edges that form a directed tree with a single source. Similarly, a \emph{backward-subtree} of $G$ is a subset of edges that form a directed tree with a single target.
%\footnote{These are called \emph{out-branching} and \emph{in-branching}, respectively, in \cite{kreutzer2016nowhere}.}

A directed 2-terminal graph with no cycles is referred to as \emph{TDAG} (2-Terminal Directed Acyclic Graph). The vertices of a TDAG can always be sorted in increasing order, called \emph{topological order}, so that all edges, and thus all directed paths, are from $v_i$ to $v_j$ for some $j>i$. In particular, $s$ and $t$ are the first and last vertices, respectively. Also note that in a TDAG all paths are simple. 
\noijcai{
For technical reasons we also allow the degenerated TDAG with a single vertex (serving as both source and target). } 

We say that a vertex of a directed graph is a source if it has no incoming edges and at least one outgoing edge. A vertex is a sink in a directed graph if has no outgoing edges and at least one incoming edge.

\begin{lemma}
\label{lem:tdag-char}
A DAG is TDAG if and only if it has a unique source and a unique sink.
\end{lemma}
\begin{proof}
Clearly, if a DAG has more than one source, then at most one can be $s$ and thus there
are no $s-t$ paths that visit the rest of the sources. Similarly for sinks and vertex $t$.
Now, in order to prove the other direction, i.e., that a DAG with a unique source and a unique
sink is a TDAG, we need to prove that for every edge of this graph there exists a simple 
$s-t$ path. This is not hard to see. Consider an arbitrary edge $(v_i,v_j)$ of the DAG and let
$s, v_1, \ldots, v_n, t$ be a topological order of the DAG. Take the two subgraphs induced
by $s, v_1, \ldots, v_i$ and $v_j, \ldots, v_n, t$. These graphs are DAGs where $v_i$ is a
sink in the first one and $v_j$ is a source in the second one. So, there exists a
simple path $[p_{sv_i}]$ in the first induced subgraph and  a simple path $[p_{v_jt}]$ in 
the second induced subgraph. So, in the original graph there exists the $s-t$ path 
$[p_{sv_i}]-[p_{v_jt}]$.
\end{proof}

In undirected graphs, a graph $H'$ is called a \emph{minor} of $H$ if  $H'$ can be obtained from $H$ by a sequence of edge deletions and contractions. 
The contraction of an edge $(a,b)$ creates a new node instead of $a$ and $b$, 
whose set of neighbors is the union of $a$'s and $b$'s neighbors.
As an example of a simple characterization via exclusion of minors, observe that any graph  $H$ (not a multigraph) is acyclic if and only if it excludes a triangle as a minor. 
%\adl{
%We extend the following definition of minors in undirected graphs to undirected graphs.
%
%\begin{definition}[Graph minor] \label{def:minor}
%A  graph $H'$ is a \emph{minor} of undirected graph $H$, 
%if $H'$ can be obtained from $H$ by a sequence of the following  operations:
%\begin{description}[topsep=0.5ex,itemsep=0.1ex,labelindent=1em]
%	\item[Deletion.] Deleting an edge $(a,b)$;
%	\item[Contraction.] Contracting an edge $(a,b)$.
%\end{description}
%Contraction creates a new node instead of $a$ and $b$, whose set of neighbors is the union of $a$'s and $b$'s neighbors.
%\end{definition}
%As an example of a simple characterization via exclusion of minors, observe that  graph  $H$ (not a multigraph) is acyclic if and only if it excludes a triangle as a minor. This is since if $H$ is cyclic then we can delete all edge that do not belong to the cycle, and then contract all edges of the cycle except three. However if the graph is acyclic it remains so after any minor operation so we will never get a triangle (which is cyclic) as a minor. 
%\rmr{add example}}

%%%%%%++++++++++++++
%%%%%%++++++++++++++

\section{Directed Graph Minors and Embeddings} \label{sec:minors}
\subsection{Directed minors}
%\paragraph{\bf Directed minors.}
%An undirected graph $G'$ is a \emph{minor} of graph $G$, if $G'$ can be obtained from $G$ by contracting edges, deleting edges, and deleting isolated vertices (see e.g.~\cite{lovasz2006graph}). 
There are several extensions of the notion of a minor to directed graphs. One that is closest to our needs is the \emph{butterfly minor}~\cite{johnson2001directed}, see Def.~\ref{def:dminor} without the underlined part. However, neither the class of 2-terminal graphs nor the class of  TDAGs is closed under the butterfly minor operation, since, for example, it may leave an isolated node. We thus  modify it by restricting which edges may be deleted (underlined).

\begin{definition}[Directed minor]\label{def:dminor}
A  graph $G'$ is a \emph{directed minor} (or simply \emph{a d-minor}) of a  directed graph $G$, 
 if $G'$ can be obtained from $G$ by a sequence of the following local operations:
\begin{description}[topsep=0.5ex,itemsep=0.1ex,labelindent=1em]
	\item[Deletion.] Deleting an edge $(a,b)$ \underline{where $a$ has outdegree at least~2, and $b$ has indegree at least~2}.
	\item[Backward contraction.] Contracting an edge $(a,b)$ where $b$ has indegree~1.
	\item[Forward contraction.] Contracting an edge $(a,b)$ where $a$ has outdegree~1. 
\end{description}
 \end{definition}
%As with d-embeddings, it is easy to see that a d-minor inherits the property of being 2-terminal (albeit d-minors are well-defined for any directed graph), since every non-contracted edge is still part of a simple $s-t$ path.

Observe that for any  $e\in E$ of a directed graph $G=\tup{V,E}$, there is a d-minor step that either deletes or contracts $e$.
This is simply since every edge meets the premise of one d-minor operation. 
For example, the edge $(a,b)$ in Fig.~\ref{sfig:Braess} may not be contracted, but can be backward-contracted after the edge $(s,b)$ is deleted. 

The class of TDAGs is closed under d-minor operations. So, if $G$ is TDAG and $G'$ is a d-minor of $G$, then $G'$ is a TDAG. 

\begin{lemma}
\label{lem:dminor-closed}
The class of directed acyclic graphs are closed under d-minor operations.
\end{lemma}

\begin{proof}
Let $G'=\tup{V',E'}$ denote a d-minor of $G$.
Acyclicity is clearly maintained after edge deletion. Contraction steps create  no new paths (and in particular no cycles). Suppose that graph $G'$ after the backward contraction of edge $(a,b)$ has a path $p'=x-a-y$ for some pair $x,y$. Then since before contraction $b$ had indegree~1, the only incoming edge to $b$ was from $a$, which means that path  $[p'_{xa}]-b$ exists in $G$. Also, either the path $[p'_{by}]$ or $[p'_{ay}]$ is in $G$, and thus $x,y$ are connected in $G$ via the path $[p'_{xa}]-p'_{ay}-y$ or $[p'_{xa}]-[p'_{by}]$. The proof for forward induction is similar.

We next show that $G'$ remains 2-terminal graph. Consider first a backward contraction step of edge $(a,b)$. For every $s-t$ path $p$ that goes through $(a,b)$ in $G$, the simple path $s-p_{sa}-a-p_{at}-t$ exists in $G'$, thus $G'$ is a valid 2-terminal graph. 
Consider first a deletion step of edge $(a,b)$. We need to show that every edge $e\in E'$ is on some $s-t$ path in $G'$. Let $p$ be some $s-t$ path in $G$ containing $e$. If $(a,b)$ is not in $p$ then $e$ is not affected. If $(a,b)$ in $p$, consider the subpaths $p_{sa}$ and $p_{bt}$ of $p$. One of them contains $e$, w.l.o.g. $p_{sa}$. Note that since $a$ has outdegree at least $2$ in $G$, there is an edge $(a,a')$ in $E'$. Moreover, this edge must belong to some simple path $p'=[p_{sa}]-a'-t$  in $G$, and this path may not contain $(a,b)$ as this would mean that $a$ appears twice in $p'$. % Let $p'_a$ be the subpath $a-a'-t$ of path $p'$. 
The concatenation of $p_{sa}$ and $p'_{at}$ is a $s-t$ path that contains $e$ in $G'$.
\adl{do we have to define what indegree, outdegreeare?}
%
%Next, consider a backward contraction step of edge $(a,b)$. Every $s-t$ path $p$ that goes through $(a,b)$ in $G$, the path $s-p_{sa}-a-p_{at}-t$ exists in $G'$, thus $G'$ is a valid 2-terminal graph. We also note that there are no new paths: suppose that $G'$ has a path $p'=x-a-y$ for some pair $x,y$. Then since before contraction $b$ had indegree~1, the only incoming edge to $b$ was from $a$, which means that path  $[p'_{xa}]-b$ exists in $G$. Also, either the path $[p'_{by}]$ or $[p'_{ay}]$ is in $G$, and thus $x,y$ are connected in $G$ via the path $[p'_{xa}]-p'_{ay}-y$ or $[p'_{xa}]-[p'_{by}]$.
\end{proof}

\rmr{It is an interesting question what other properties of minor graphs extend to the directed minor graphs, but it is outside the scope of the current paper.}

\subsection{Graph Embeddings}\label{sec:embed}
%\paragraph{\bf Graph Embeddings.}
There are various definitions of graph embeddings and subdivisions~ \cite{ForHopWyl80,milchtaich2006network,holzman2003network},
which can be summarised together as follows.
\begin{definition}[Homeomorphic embedding]\label{def:h_embed}
A [directed]  graph $G'$ is \emph{h-embedded}  in $G$% if $G'$ is homeomorphic  to a subgraph of $G$. That is
, if $G$ (or a graph isomorphic to $G$) can be derived from $G'$ by a sequence of the following  operations:
\begin{description}[topsep=0.5ex,itemsep=0.1ex,labelindent=1em]
	\item[Addition.] The addition of a new edge joining two existing vertices.
	\item[Subdivision.] Replacement of an edge $(a,b)$ by two edges $(a,x)$ and $(x,b)$.
	\item[Terminal extension.] (Only for 2-terminal graphs.)  Addition of a new edge $e$ joining $s$ or $t$ with a new vertex, which becomes the new source or target.
\end{description}
\end{definition}
%
%A \emph{subgraph} of $G$ is a graph $G'$ obtained only by removing edges and vertices. An easy observation is that if $G'$ is a valid subgraph of $G$ (i.e., $G'$ is also a 2-terminal graph), then $G'$ is h-embedded in $G$. 

For an undirected graph $H'$,  every h-embedding operation maintains various properties like being a 2-terminal graph. % Thus the notion of embedding captures the idea that $H'$ is ``simpler'' than $H$. 
However, for a 2-terminal directed graph $G'$, an h-embedding operation may not maintain this property (see Fig.~\ref{sfig:no_directed_embed}).  Also, this set of operations is not rich enough for our needs. Thus, we  propose a new definition for directed embeddings. %, which we note that is not restricted to 2-terminal graphs. 

\begin{center}
\begin{figure}
\scalebox{0.7}{
%\begin{minipage}[b]{0.4\textwidth}
%\label{sfig:no_directed_embed}
\input{no_directed_embed}
%\end{minipage}
}
~~~~~~~~~~~~~~
\begin{minipage}[b]{0.4\textwidth}
\scalebox{0.7}{
\input{no_subdivide}
}
\end{minipage}
\small{\caption{\label{fig:examples}Examples.
The graph in Fig.~\ref{sfig:no_directed_embed} is directed 2-terminal graph  (solid edges only). Adding the dashed edge $(x,y)$, regardless of its direction,  results in a non-2-terminal graph. Fig.~\ref{sfig:subdivide}: The graph $G'$ on the left is d-embedded in $G$ on the right, as we can forward-split $u$ into $(u,v)$ ($u$ retains all incoming edges, and $v$ retains at least one outgoing edge). However, there is no edge we can add or subdivide to get $G$ from $G'$ so $G'$ is not h-embedded in $G$. The Braess graph $G_B$ is on Fig.~\ref{sfig:Braess}.
}}
\end{figure}
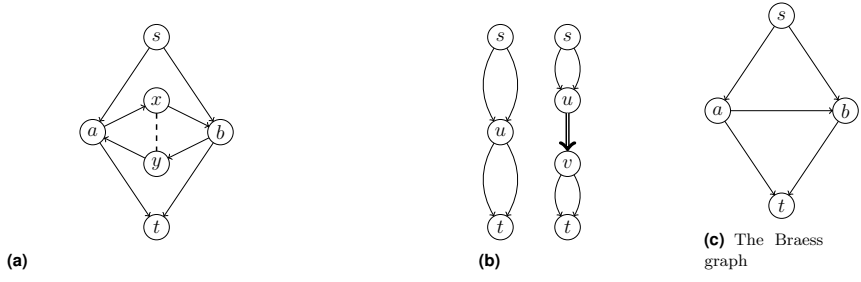
\end{center}

%\vspace{-3mm}
\begin{definition}[Directed embedding]\label{def:d_embed}
A directed graph $G'$ is \emph{d-embedded}  in a directed graph $G$ if $G'$ is isomorphic to $G$ or to a graph
derived from $G$  by a sequence of the following  operations:
\begin{description}[topsep=0.5ex,itemsep=0.1ex,labelindent=1em]
\item[Addition.] Addition of a new edge $(a,b)$, such that there is no path $b-a$.
	\item[Forward split.] Replacement of node $a\neq t$ by two nodes and an edge $(a,b)$, where $a$ retains all incoming edges, and $b$ retains at least one outgoing edge.
	\item [Backward split.] Replacement of node $b\neq s$ by two nodes and an edge $(a,b)$, where $b$ retains all outgoing edges, and $a$ retains at least one incoming edge.
	%\item The extension of a terminal vertex: addition of a new edge $e$ joining $s$ or $t$ with another, new vertex, which becomes the new source or target, respectively;
\end{description}
\end{definition}

Note that if $G'=\tup{V',E'}$ is d-embedded in $G=\tup{V,E}$, then we can identify for each node $x\in V'$ a corresponding node in $V$. This mapping may not be unique, as there may be several ways to obtain $G$ from $G'$. However the mapping is unique for a given sequence of d-embedding operations. 

%\begin{remark}\label{rem:split}
It is not hard to see that a subdivision of an edge (directed or undirected) can be replicated by splitting one of its end nodes, and a terminal extension can be replicated by splitting the terminal (backward split of $s$ or forward split of $t$).  We thus allow the operations of \textbf{edge subdivision} and \textbf{terminal extension} as valid d-embedding operations as well.
%\end{remark}
%
%The new definition maintains important graph properties. One of them guarantees that the class of 
%2-terminal directed graphs is closed under d-embedding. 
%So, if $G'$ is a TDAG and $G'$ is d-embedded in $G$, then $G$ is a TDAG.

\begin{lemma}\label{lemma:embed_closed}
The classes of 2-terminal directed graphs and directed acyclic graphs are closed under d-embedding.
\end{lemma}
In particular, if $G'$ is a TDAG and $G'$ is d-embedded in $G$, then $G$ is a TDAG.

\begin{proof}
Suppose first that $G'$ is acyclic. By definition of d-embedding, adding an edge maintains acyclicity. Split operations do not add any new paths and in particular cannot add cycles. 

Next, suppose that $G'$ is 2-terminal. We add an edge $(a,b)$. 
 Clearly all existing edges and vertices are still part of some $s-t$ path. Consider some paths $p'=s-a$ and $p''=b-t$. 
Since there is no path from $b$ to $a$ in $G'$, the paths $p',p''$ do not intersect.  Thus,
the new edge $(a,b)$ is part of the simple path $p=s-p'-a-b-p''-t$ in $G$. %This path must be simple since $G$ is a DAG. 
Finally, consider  a (forward) split step of vertex $a$, the path $p'=s-a-x-t$ becomes the path $p=[p'_{sa}]-b-[p'_{xt}]$, %where $x$ is the node following $a$ in $p'$, 
and $p$ contains $(a,b)$. 
\end{proof}

%A subgraph $G'$ of a directed 2-terminal graph $G$ is a graph obtained from $G$ by deleting edges and vertices that are not $s$ or $t$, such that $G'$ is still 2-terminal.
%A subgraph $G'$ of a 2-terminal graph $G$ is a graph obtained from $G$ by deleting edges and vertices that are not $s$ or $t$, or by deleting $s$ or $t$ together with their unique connecting edge. 

  %If $G'$ is a subgraph of $G$ we say that $G$ is a \emph{supergraph} of $G'$. 
For a 2-terminal directed graph $G$, the graph $G'$ is a \emph{valid subgraph} of $G$ if it is 2-terminal and isomorphic to a subgraph of $G$. While the next lemma may seem trivial, note that it does not hold for general 2-terminal directed graphs, since
a single edge is not d-embedded in any cyclic graph. 

% \emph{Deleting a path} $a-b$ from $G$ is possible if all intermediate vertices have indegree and outdegree~1, and means deleting all intermediate vertices and edges of the path $a-b$.
\begin{lemma}\label{lemma:subgraph}
If $G'$ is a valid subgraph of TDAG $G$, then $G'$ is d-embedded in $G$.% it is (isomorphic to a graph) obtained from $G$ by iteratively deleting paths. 
\adl{define valid subgraph? Proof?}\rmr{define above}
\end{lemma} 

\begin{proof}
We prove by induction on the difference $|E|-|E'|$. Let $e=(a,b)$ be an edge in $E\setminus E'$. As $e$ is part of  path   $p=s-a-b-t$ in $G$, let $x$ be the last node on the subpath $[p_{sa}]$ that is also in $V'$ (such a node must exist since $s\in V'$). Also let $y$ be the first node on $[p_{bt}]$ that is in $V'$. Let $k$ be the length of the path $[p_{xy}]$. We perform the following d-embedding steps: add an edge $(x,y)$ to $G'$, and subdivide it $k-1$ times. The result $G''$ is a valid subgraph of $G$, that has strictly more than $|E'|$ edges. Thus by induction there is a sequence of d-embedding steps from $G''$ to $G$, and hence from $G'$ to $G$.
\end{proof}

We will need the following lemma later on, but it is useful to know regardless. An immediate corollary is that embedding steps only add paths and increase the connectivity of a graph. 
\begin{lemma}\label{lemma:split_path}
If $G,G'$ differ by a single  split step of vertex $a$ into $(a,b)$, then there is a one to one mapping between paths in $G'$ to paths in $G$. 
\end{lemma}

\begin{proof}
For any path $p'=x-y$ in $G'$ there is a path $p$ in $G$ as follows:
\begin{itemize}[topsep=0.5ex,itemsep=0.1ex,labelindent=1em]
	\item If $a\notin p'$ then $p=p'$.
	\item If $p'=x-a-y$ then either $p=p'$ or $p=[p'_{xa}]-[p'_{by}]$.
\end{itemize}
For any path $p=x-y$ in $G$ where $x \neq b$, it is not possible that $p$ contains $b$
but not $a$. So, for any path $p=x-y$ there is a path $p'$ in $G'$ as follows:
\begin{itemize}[topsep=0.5ex,itemsep=0.1ex,labelindent=1em]
	\item If $b\notin p$ then $p'=p$.
	%\item If $p=x-a-y$ and $b\notin p$ then $p' = x-a-y$;
	\item If $p=x-a-b-y$ then $p'=x-p_{xa}-a-p_{ay}-y$.
\end{itemize}
\end{proof}

%Backward contraction of $(a,b)$ eliminates the node $b$, whereas forward contraction eliminates $a$. Thus we can identify every node $v$ in $G'$ with a single original node $Orig_G(v)$ in $G$.
%We argue that none of the steps above, and in particular contraction, introduces new paths. In particular this means that d-minors inherit the property of being cycle-free.
%
%We get the next lemma as an immediate corollary, as addition steps never eliminate paths.
%\begin{lemma}\label{lemma:embed_path}
%If $G'$ is  d-embedded in $G$, and $G'$ contains a  path from $x$ to $y$, then $G$ contains a path from $x$ to $y$.  
%\end{lemma}

%\paragraph{Graph subdivision}

\subsection{Relations among graph operations}
The way we defined them, d-minors are more restrictive than butterfly minors, whereas d-embeddings are more permissive than h-embeddings; see Fig.~\ref{sfig:subdivide}. 
%By Remark~\ref{rem:split}, directed embedding are strictly more permissive than homeomorphic embeddings. 
However, d-embeddings are not infinitely richer than h-embeddings. A vertex is called a \emph{hub} if it  has both an indegree and an outdegree larger than one.

\begin{lemma}\label{lemma:subdivide}
Let $G,G'$ be TDAGs. If $G$ is h-embedded in $G'$, then $G'$ is d-embedded in $G$. 
\end{lemma}
\begin{proof}
We first recall that subdivision of an edge is a valid d-embedding step. 
H-embeddings may allow adding edges the close a cycle, but since $G$ is a TDAG it does not require the addition of such edges.
 %
%Denote the supergraph of $G'$ by $\hat G$. Thus $G'$ is d-embedded in $\hat G$, and by Lemma~\ref{lemma:embed_closed}, $\hat G$ is also a TDAG. Since $\hat G$ is a TDAG isomorphic to a subgraph of $G$, then by Lemma~\ref{lemma:subgraph} $\hat G$ is d-embedded in $G$, and thus $G'$ is d-embedded in $G$. 
\end{proof}
The converse of Lemma~\ref{lemma:subdivide} does not hold. That is, d-embedding allows for a richer set of operations than subdivision and adding edges. 
To see this, consider the graphs in Fig.~\ref{sfig:subdivide}.

\def\calG{\mathcal{G}}
\begin{proposition}\label{prop:embed_finite}
Let $G'=\tup{V',E'}$ and $G=\tup{V,E}$ be graphs such that $G'$ is d-embedded in $G$. Let $J\subseteq V'$ be the hubs of $G'$.
There is a set $\calG$ of  at most $2^{|J|\times |V'|^{2}}$ graphs, such that some graph in $\calG$ is h-embedded in $G$. Each such graph has at most $|V|(1+|J|)$ vertices.   
\end{proposition}

\begin{proof}
Consider a sequence of d-embedding steps from $G'$ to $G$. 
  Any split operation on a non-hub is equivalent to an edge subdivision. W.l.o.g. all the last operations in the sequence are edge additions (by Lemma~\ref{lemma:subgraph}), and before them all the subdivision operations. Thus, all the split operations on hubs are w.l.o.g. the first $k'\leq |J|\times|V'|$ steps of the sequence. Note that the graph we obtain after $k'$ steps is h-embedded in $G$. We construct the set $\calG$ by performing every  possible combination of split operations on $G'$. This means selecting a subset of hubs, and split each selected hub to a tree that contains no hubs. There are clearly no more that $2^{|V'|^2}$ such trees, thus $|\calG| \leq \prod_{j\in J}2^{|V'|^2} =2^{|J|\times |V'|^{2}}$.  Since each split operation reduces either the indegree or the outdegree of exactly one hub by 1, the total size of the expanded graph is $|V|+\sum_{j\in J}(outdeg(j)+indeg(j)-2)\leq |V| + \sum_{j\in J}|V| = |V|(1+|J|)$ \adl{def of $deg(j)$? Incoming, outgoing, combined?}.
\end{proof}

%%%%%%%%%%%%%%%%%%%%%%%%%%%%%%%%%%%%%%%%%%%%%

For the class of TDAGs, the concepts of directed-minor and directed-embedding turn out to be equivalent.
\begin{theorem}\label{thm:minor_embed}
Let $G$ and $G'$ be TDAGs.   $G'$ is d-embedded in $G$ if and only if $G'$ is a d-minor of $G$. %\footnote{A similar equivalence exists between graph subdivisions and butterfly minors.}
\end{theorem}
\adl{proof?}

Intuitively, addition and deletion operations cancel one another, as do split and contraction operations. 
This equivalence does not hold for general directed graphs, as added edges may not qualify for deletion (e.g. if we add an edge $(a,b)$ where $a$ has only incoming edges), and vice versa (if we remove an edge that is part of a cycle).  
\begin{proof} By induction, it is sufficient to show this for $G',G$ that differ by a single d-embedding or d-minor operation. 
``$\Rightarrow$'' %Consider a sequence of graphs $G'=G^0,G^1,\ldots,G^m=G$, where each graph differs from the previous one by a single embedding operation. It is sufficient to show that each $G^i$ is a d-minor of $G^{i+1}$. 
There are 3 cases, depending on the embedding operation:
\begin{enumerate}
	\item  The addition of  edge $(a,b)$ to $G'$ can be reversed by deleting the same edge from $G$. Note that $b\neq s$ as otherwise there is a path in $G'$ from $b=s$ to $a$, and similarly $a\neq t$. Thus $a$ has outdegree at least 1 in $G'$ and at least 2 in $G$. Similarly, $b$ has indegree at least 2 in $G$, and thus deleting the edge $(a,b)$ is a valid d-minor step.
	\item Suppose that a vertex $a$  in $G'$ is split to $\{a,b\}$ with a forward split. Then since $a$ retains all incoming edges, $b$ has a single incoming edge $(a,b)$ in $G$. Thus we can contract the edge $(a,b)$ in $G$ using backward contraction.
	\item Similarly, a backward split can be reversed with a forward contraction.%Suppose that a vertex $b$ is split to $\{a,b\}$ with a backward split. Then since $b$ retains all outgoing edges, $a$ has a single outgoing edge $(a,b)$. Thus we can contract the edge $(a,b)$ using forward contraction.
	%\item  The extension of a terminal vertex $s$ in $G'$ by a new terminal $s'$ and edge $(s',s)$ can be reversed by deleting $(s,s')$ and then $s'$. Similarly for $t$. 
\end{enumerate}

``$\Leftarrow$'' %Consider a sequence of directed 2-terminal graphs $G=G^0,G^1,\ldots,G^m=G'$, where each graph differs from the previous one by a single d-minor operation. It is sufficient to show that each $G'$ is  d-embedded in $G'$. 
There are 3 cases, depending on the d-minor operation:
\begin{enumerate}
	\item If the edge $(a,b)$ is deleted from $G$, then since $G$ is acyclic there is no path $b-a$. Thus adding $(a,b)$ to $G'$ is a valid d-embedding step.
	\item Suppose that the edge $(a,b)$ in $G$ is backward-contracted to some vertex $x$ in $G'$. This means that $b$ has a single incoming edge. Thus all edges incoming to the pair $\{a,b\}$ are leading to $a$. Let $R(a), R(b)$ be the out-neighbors of $a$ and $b$ in $G$, respectively. Then by forward-splitting node $x$ in $G'$ and split the outgoing edges of $x$ according to $R(a)$ and $R(b)$, we get the graph $G^{i}$. 
	\item Similarly, forward contraction can be reversed with backward split. 
%	\item If $G'$ is a subgraph of $G$ then by Lemma~\ref{lemma:embed_subgraph} $G'$ is d-embedded in $G$.
	%\item  The extension of a terminal vertex $s$ in $G'$ by a new terminal $s'$ and edge $(s',s)$ can be reversed by deleting $(s,s')$ and then $s'$. Similarly for $t$. 
\end{enumerate}
\end{proof}

%Can we extend the characterization to more general graphs?
\noijcai{
\rmr{keep?}
A major question is whether a result similar to Wagner's conjecture can be proven for directed graphs, or at least for TDAGs.
\begin{conjecture}[A Wagner conjecture for TDAGs]\label{conj:d_wagner}
For every class $\mathcal {G}$ of TDAGs that is closed for d-minor operations, there is a finite set of TDAGs $R=\{G_1,\ldots,G_r\}$ such that $G\in \mathcal {G}$ if and only if no $G_i\in R$ is a d-minor of $G$. 
\end{conjecture}}

\section{Serial-Parallel Width}\label{sec:SPW}
A \emph{cut} in a 2-terminal graph $G = \tup{V,E,s,t}$ is a set of edges $C\subseteq E$ such that there is
no $s-t$ path in $E\setminus C$. $C$ is \emph{minimal} if there is no cut $C'\subsetneq C$.  
%The following 3 definitions are due to \cite{MP18}.
%\begin{definition}\label{def:parallel}%\label{def:par_edges}

A set of edges $S\subseteq E$  is \emph{parallel} if there is some $C \subseteq E$ s.t. $S\subseteq C$, and $C$ is a minimal cut between $s$ and $t$ in the graph $G$; $S$ is \emph{serial} if there is a simple directed $s-t$ path $p$ that contains $S$.    %\footnote{\label{fn:parallel}We can further restrict the definitions to a subset of valid $(s_i,t_i)$ pairs (e.g. all source-target pairs in a given game $\GG$). This may only reduce the parallel width and thus improve our results.}
%	\end{definition}
	%
%
\def\SPW{SPW}
\def\PW{PW}
%\begin{definition}\label{def:serial_edges}
% A set of edges $S\subseteq E$ 

%\end{definition}

\begin{definition}[Parallel Width]\label{def:PW}
The \emph{parallel width} of a directed 2-terminal graph, $\PW(G)$, is the size of the largest parallel set $S\subseteq E$. 
\end{definition}

\begin{definition}[Serial-Parallel Width~\cite{MP18}]\label{def:SPW}
The \emph{serial-parallel width} of a directed 2-terminal graph, $\SPW(G)$, is the size of the largest set $S\subseteq E$ that is both serial and parallel. 
\end{definition}

Intuitively, the parallel width is the size of a maximum $s-t$ cut. For example, the width of an electric circuit coincides with the parallel width of its underlying TDAG~\cite{codenotti1991parallel}.  A serial-parallel width of $k$ means that there are at least $k$ source-target paths, and some additional path that edge-intersects all of them.  It was shown in \cite{MP18} that in nonatomic routing games with diverse players, the negative externality is bounded  by the serial-parallel width of the underlying network.
In the context of routing games, a serial-parallel width of 2 means that there is a group of agents that can negatively influence two other groups~\cite{MP18}. 

\begin{example}
Consider the Braess graph  in Fig.~\ref{sfig:Braess}. The minimal $s-t$ cuts are: $\{sa,sb\}$,  $\{at,bt\}$, $\{sa,bt\}$, and $\{sb,ab,at\}$. Thus, the set $\{sa,bt\}$ is both parallel and serial, which means $\SPW(G_B)\geq 2$. The set $\{sa,at\}$ is serial but not parallel; and  $\{sa,sb,ab\}$ is neither. In fact, the \emph{only} parallel set of size greater than $2$ is $\{sb,ab,at\}$, which is not serial, thus $\SPW(G_B)<3$. We conclude that the serial-parallel width of the Braess graph is $2$.
In contrast, both graphs in Fig.~\ref{sfig:subdivide} have $PW(G)=2$ and $SPW(G)=1$.
\end{example}

\noijcai{ 
\rmr{\begin{observation}\label{ob:SPW} For any directed graph $G$, $SPW(\ol G)\geq SPW(G)$. 
\end{observation}
This is since any subset $S\subseteq E$ in $G$ that is serial and/or parallel, remains serial and/or parallel once we eliminate edge directions. }}
%\rmr{Better definition?: For $k\geq 2$, a directed graph $G=(V,E)$ is \emph{$k$-parallel-free} ($k$-PF) if it does not have the following structure: a minor that is a $k$-parallel graph, and a path that edge-intersects all paths of the minor.}

For any 2-terminal graph $G$, we have $1\leq \SPW(G)\leq |V|-1$. The lower bound is since any single edge is both parallel and serial, and the upper bound since there is no simple path of length $|V|$ or more.

\begin{definition}Let $G$ be a 2-terminal directed graph. A set of edges $S\subseteq E$  is \emph{concurrent} if for any $e\in S$ there is some $s-t$ path $p_e$ such that $p_e\cap S = \{e\}$.
\end{definition}
\begin{lemma}\label{lemma:concurrent} 
 If $S$ is parallel, then $S$ is concurrent.
\end{lemma}
\begin{proof}
For every $i\leq k$, there must be some path $p_i$ from $s$ to $t$ such that $e_i\in p_i$, and  $e_j\notin p_i$ for all $j\neq i$, otherwise we could drop $e_i$ from $S$ and still get a $s-t$ cut $C\setminus\{e_i\}$. 
\end{proof}
The converse does not hold. For example, the set $\{e_1,e_2\}$ in Fig.~\ref{sfig:subdivide} is concurrent but not parallel.

%
%\begin{lemma}\label{lemma:parallel_components}
%Let $G$ be a 2-terminal directed graph.  If a set of edges $S=\{(a_i,b_i)\}_{i\leq k}$ is parallel then there are two vertex-disjoint connected components $T_s,T_t$, such that $T_s$ contains $s$ and all of $a_i$, and $T_t$ contains $t$ and all of $b_i$.
%\end{lemma}
%\begin{proof}
%Suppose that $S$ is parallel, then it is contained in a minimal cut $C$. Let $G_C$ be graph $G$ without the edges of $C$. Let $T_s$ be all vertices reachable from $s$ in $G_C$, and $T_t$ all vertices from which $t$ is reachable. These are clearly disjoint as otherwise there is a path from $s$ to $t$ in $G_C$. Also, $a_i\in T_s$ for all $i$, as otherwise the edge $e_i$ can be removed from $C$ and $C\setminus\{e_i\}$ is still a cut. Likewise for $b_i\in T_t$. 
%%
%%In the other direction, consider an arbitrary minimal cut $C'$ in $G_S$. We argue that $C'\cup S$ is a minimal $s-t$ cut in $G$. Suppose we remove some edge $e=(a,b)\in S$ from the cut $C'\cup S$. Since $a\in T_s
%\end{proof}

\begin{definition}\label{def:SP_k}
For any $k\geq 2$, we define the \emph{$k$-serial-parallel graph} $G_{SP(k)}$ as follows. $G=\tup{V,E,s,t}$, where  $V=\{s,t,a_2,\ldots,a_k,b_1,\ldots,b_{k-1}\}$, and $E=\bigcup_{i=2}^{k-1}\{(s,a_i),(a_i,b_i),(b_i,t),(b_i,a_{i+1})\} \cup \{(s,b_1),(a_k,t)\}$. 
\end{definition}

\begin{definition}\label{def:SP_k_var}
A graph $G$ is a \emph{variant of $G_{SP(k)}$} if we replace the edges $\{(s,a_i)\}_{i=2}^k$ with an arbitrary forward-subtree that respects the lexicographic order $(s,a_2,\ldots,a_k)$, and replace the edges $\{(b_i,t)\}_{i=1}^{k-1}$ with an arbitrary backward-subtree that respects the lexicographic order $b_1,\ldots,b_{k-1},t$. %(note the set of vertices remains the same).
\end{definition}

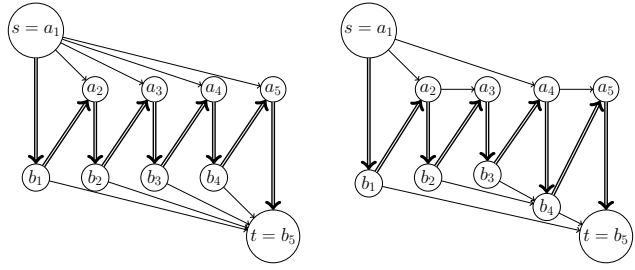
\begin{figure}
\begin{center}
\scalebox{0.65}{
\input{network_fig_SPk_variants}}
\end{center}
\small{\caption{\label{fig:SPk_var}The left figure is the graph $G_{SP(5)}$, and the right figure is a variant of it. For convenience, the long path in each graph appears in double lines, and the forward- and backward-trees in thin lines.}}
\end{figure}

See  Figure~\ref{fig:SPk_var} for examples.
The $2$-serial-parallel graph has only one variant, which is the Braess graph $G_B$. 
There are other generalizations of the Braess graph but we are unaware of one that coincides with ours.\footnote{Lianeas et al.~\cite{lianeas2015asymptotically}, for example, generalize the Braess graph to a family of graphs with long alternating paths. All these graphs have parallel-width 2, whereas $G_{SP(k)}$ graphs have alternating paths of length at most 2. }  We should note that an example of a $G_{SP(k)}$ graph appears in Nikolova and Stier-Moses~\cite{NS15} (Figure~4 there) without a formal definition or analysis, and is also used in \cite{babaioff2007congestion} to derive an example where few malicious players can hurt the others.
The serial-parallel width of $G_{SP(k)}$ is  exactly $k$, where $\{(s,b_1),(a_2,b_2),$ $\ldots,(a_{k-1},b_{k-1}),(a_k,t)\}$ are the serial-parallel edges. 

The graph $G_{SP(k)}$ was used under different names in \cite{babaioff2007congestion,NS15,MP18}, usually to derive  examples of games with high equilibrium costs.

%
%\begin{proposition}[\cite{milchtaich2006network}] For an undirected two-terminal network $G$, the following conditions are equivalent:
%(i) $G$ is series-parallel.
%(ii) For every pair of distinct vertices $u$ and $v$, if $u$ precedes $v$ in some route $r$ containing both vertices, then $u$ precedes $v$ in all such routes.
%(iii) The (undirected) Braess graph is not embedded in $G$.
 %\dcp{need to give the base case as well} 
%The basic graph is a single directed edge $s-t$. Two DSPs $(V_1,E_1,s_1,t_2), (V_2,E_2,s_2,t_2)$ can be composed to obtain a new DSP either in a {\em serial way}: merge  $s_1$ with $t_2$; or in a {\em parallel way} by merging $s_1$ with $s_2$, and $t_1$ with $t_2$. A (non-directed) series-parallel graph is obtained by taking a DSP and removing edge orientations. 

% \dcp{consider making this prev sentence a FN unless it's going to be meaningful to most readers} 

%
%While the definition was initially suggested to undirected graphs, it also applies to directed graphs~\cite{robertson2004graph}.\footnote{Note that we cannot ignore edge directions during contraction. For example, the directed graph $\{(s,t)\}$ is not a minor of $\{(s,x),(t,x)\}$, but their undirected versions do hold the minor relation.}
%\begin{rtheorem}{th:minor_SPW}
 %For a 2-terminal DAG $G$, the following conditions coincide:
%\begin{enumerate}
	%\item $SP(k)$ is d-embedded in $G$.
	%\item $SP(k)$ is a d-minor of $G$.
	%\item $\SPW(G) \geq k$.
%\end{enumerate}
 %\end{rtheorem}

\begin{lemma}\label{lemma:embed}
%If $SPW(G')\geq k$ and $G'$ is d-embedded in $G$, then $SPW(G)\geq k$. 
If $S$ is a set of parallel [serial] edges in a 2-terminal graph $G'$, then after any sequence of d-embedding steps on $G'$, the set $S$ is still parallel [resp., serial]. In particular, if $G'$ is d-embedded in $G$ then $PW(G)\geq PW(G')$ and $SPW(G)\geq SPW(G')$.
\end{lemma}
\begin{proof}
For serial sets the statement is obvious. 

%Denote graph $G'=\tup{V',E',s',t'}$. Denote by $S'$ some set of $k$ edges in $G'$ that are parallel and serial, and by $C'$ some minimal cut containing $S'$. 
Consider a sequence of $J$  d-embedding operations on $G^0=G'$ that ends in $G^J=G$. 
Suppose that $S$ is parallel.
Let $C^0$ be a minimal cut in $G^0=G'$ containing $S$. We show by induction that after every step $j\leq J$ there is a minimal cut $C^j$ in $G^j$, such that $C^{j-1}\subseteq C^j$.

 Assume by induction that $C^{j-1}$ is a minimal cut in $G^{j-1}$. The graph $G^j$ differs from $G^{j-1}$ either by a single added edge, or by a single split vertex.
  By Lemma~\ref{lemma:split_path} split steps to not change the set of paths, and thus $C^j=C^{j-1}$ is still a minimal cut. Thus suppose $G^j$ differs by an addition step of an edge $e=(a,b)$. Either  $C^{j-1}$ is still a  cut in $G^j$, or $e$ connects a node $a$ reachable from $s$ to a node $b$ with a path to $t$. In the latter case, $C^j = C^{j-1}\cup \{e\}$ is a cut. To see that $C^j$ is minimal suppose we remove an edge $e'\neq e$. %If we remove edge $e$ then we already know there is a path $s-a-b-t$. 
	If $C'=C^j\setminus \{e'\}$  is a cut in $G^j$, then $C'\setminus \{e\} = C^{j-1}\setminus\{e'\}$ is a cut in $G^{j-1}$, in contradiction to the induction hypothesis that $C^{j-1}$ is minimal. 
	In either case, $S$ is still contained in a minimal cut $C^j$ after every operation, and in particular contained in a minimal cut $C^J$ of $G^J=G$.
%
%
%
%It is sufficient to show that $SPW(G)\geq k$ for $G$ obtained after a single embedding operation (add edge or split node). %By Lemma~\ref{##} the set $S'$ is still crossed.
%\begin{itemize}
	%\item add edge $e$:  Either $C'$ is still a minimal cut, or $C'\cup \{e\}$ is. Also, the path containing $S'$ still exists after adding an edge.
	%\item split node: By Lemma~\ref{lemma:split_path}, $G$ and $G'$ have exactly the same set of paths, and thus $C'$ is still a minimal cut in $G$.  For the same reason, there is a  path containing $S'$ in $G$.
	%%
	%%\item subdivide edge: if we subdivide an edge in $C'$, select one of the new edges arbitrarily. $C'$ is still a minimal $s'-t'$ cut in the new graph, and thus $S'$ is contained in a minimal cut.
	%%\item extend terminal: if we extend $s'$ then we denote the new terminal by $s$ (and add an edge $(s,s')$). Similarly, if we extend $t'$ then the new edge is $(t',t)$. After any such extension, $C'$ is still a minimal $s'-t'$ cut in the new graph. 
%\end{itemize}
%In either case we get that the set $S'$ is both parallel and serial in the new graph, so $SPW(G)\geq k$.
\end{proof}

\subsection{Characterization of graphs with bounded serial-parallel width}
Before we get to our main theorem we start with a characterization of parallel sets.
%
%\begin{proposition}\label{prop:PW_embed}
%Let $G$ be a TDAG, and let $k\geq 2$. Then $\PW(G)\geq  k$ iff  $G_{P(k)}$ is a d-minor of $G$ (or, equivalently, embedded in $G$).
%\end{proposition}
%\begin{proof}
%Suppose that $G_{P(k)}$ is d-embedded in $G$. Denote by $S$ the set of edges of $G_{P(k)}$. We denote by $C$ some minimal cut that contains $S$. Initially $C=S$.  By Lemma~\ref{lemma:split_path} split steps to not change the set of paths, and thus $C$ is still a minimal cut. When we add an edge $e$, either $C$ is still a (minimal) cut, or $e=(a,b)$ connects a node $a$ reachable from $s$ to a node $b$ with a path to $t$. Thus $C\cup \{e\}$ is a cut, which is clearly minimal. In either case, $S$ is still contained in a minimal cut $C$ after every operation.
%
%In the other direction, if there is a parallel set $S$ of size $k$, then  by Lemma~\ref{lemma:parallel_components} there are components $T_s$ and $T_t$ that connect $s$ to $S$ and $S$ to $t$, respectively. Since $G$ is a TDAG, $T_s$ is a forward-subtree and $T_t$ is a backward subtree. The graph $G' = T_s \cup S   \cup T_t$ is a valid subgraph of $G$ and thus a minor. Since all nodes in $T_s$ have indegree at most 1, we can backward-contract all of $T_s$ to a single node $s$. Similarly, we forward-contract all of $T_t$ to the node $t$, and we are left with a graph that has two nodes and $k$ parallel links, i.e. $G_{P(k)}$.   
%\end{proof}

\begin{proposition}[Parallel sets characterization]\label{prop:par_sets}
Let $G=\tup{V,E,s,t}$ be a TDAG, and a set of $k$ edges $S\subseteq E$, where for each $e_i\in S$, $e_i=(a_i,b_i)$. The following conditions are equivalent:
\begin{enumerate}
	\item $S$ is parallel;
	%\item There is a valid subgraph $G'$ of $G$ of which $S$ is a minimal cut;
	\item There is a forward-subtree $T_s$ in $G$ with root $s$ and leafs $\{a_i\}_{i\leq k}$, and a backward-subtree $T_t$ in $G$ with leaf $t$ and roots $\{b_i\}_{i\leq k}$;
	\item There is a sequence of d-minor operations that deletes or contracts all edges except $S$.
\end{enumerate}
\end{proposition}
%\noijcai{
\begin{proof}
``$1 \Rightarrow 2$'': % By Lemma~\ref{lemma:parallel_components}, there are such connected components $T_s$ and $T_t$. Since $G$ is a TDAG, and $T_s$ contains a path from $s$ to every $a_i$, then $T_s$ is w.l.o.g. a forward-tree. Similarly for $T_t$. 
	Suppose that $S$ is parallel, then it is contained in a minimal cut $C$. Let $G_C$ be graph $G$ without the edges of $C$. Let $T_s$ be all vertices reachable from $s$ in $G_C$, and $T_t$ all vertices from which $t$ is reachable. These are clearly disjoint as otherwise there is a path from $s$ to $t$ in $G_C$. Also, $a_i\in T_s$ for all $i$, as otherwise the edge $e_i$ can be removed from $C$ and $C\setminus\{e_i\}$ is still a cut. Likewise for $b_i\in T_t$. Since $G$ is a TDAG, and $T_s$ contains a path from $s$ to every $a_i$, then $T_s$ is w.l.o.g. a forward-tree. Similarly for $T_t$.
	
``$2 \Rightarrow 3$'':  The union of $T_s,S$, and $T_t$ is a valid subgraph $G'$ of $G$ of which $S$ is a minimal cut: for any $e_i$ there is a path $s-a_i-b_i-t$. 
	Since $G'$ is a valid subgraph of $G$, then by Lemma~\ref{lemma:subgraph} it is d-embedded and thus a d-minor of $G$. Then, since all nodes in $T_s$ have indegree at most 1, we can backward-contract all of $T_s$ to a single node $s$. Similarly, we forward-contract all of $T_t$ to the node $t$, and we are left with a graph that has two nodes whose only edges are $S$. %, i.e. $G_{P(k)}$. 
	
		``$3 \Rightarrow 1$'':  By Theorem~\ref{thm:minor_embed} we can consider the reverse sequence of d-embedding operations from $G^0=G_{P(k)}$ to $G^J=G$. By Lemma~\ref{lemma:embed}, the set $S$ is still parallel after every operation and in particular in $G$. 
	\end{proof}
	 %}
	We get a characterization of graphs  with bounded parallel width as a simple corollary. 
	\begin{theorem}\label{th:PW_embed}
	For any TDAG $G$ and $k\geq 2$,  $PW(G)\geq k$ if and only if  $G_{P(k)}$ is a d-minor of $G$. 
	\end{theorem}
	\begin{proof}
	``$\Rightarrow$'': Consider some parallel set $S$ of size $k$. By Prop.~\ref{prop:par_sets} there is a sequence of d-minor operations that ends with a graph whose only edges are $S$. This graph is $G_{P(k)}$.  
	``$\Leftarrow$'': Follows directly from Lemma~\ref{lemma:embed} and Thm.~\ref{thm:minor_embed}, since $PW(G_{P(k)})=k$.
	\end{proof}

\begin{theorem}[Main Theorem]\label{th:SPW_embed}
For any TDAG $G$ and $k\geq 2$,  $SPW(G)\geq k$ if and only if some variant of $G_{SP(k)}$ is a d-minor of $G$. 
%Let $G$ be a TDAG, and let $k\geq 2$. The following conditions coincide.
%\begin{enumerate}[topsep=0.5ex,itemsep=0.1ex,labelindent=1em]
	%\item $\SPW(G)\geq  k$;
	%\item Some variant of $G_{SP(k)}$ is a d-minor of $G$;
	%\item Some variant of $G_{SP(k)}$ is d-embedded in $G$.
%\end{enumerate}
\end{theorem} 
\begin{proof}
%Suppose $G$ has the Braess graph with vertices $\{s,a,b,t\}$ as a minor (where the long path is $s-a-b-t$).  Let $E_1\subseteq E$  be some minimal $s-a$ cut in $G$, and $e_1\in E_1$. Let $E_2\subseteq E$  be some minimal $b-t$ cut in $G$, and $e_2\in E_2$. The set $E_1\cup E_2$ is a minimal $s-t$ cut in a subgraph of $G$, and thus $S=\{e_1,e_2\}\subseteq E_1\cup E_2$ is is contained in some minimal cut of $G$.    Also there is a path $s-a-b-t$ in $G$ that contains $S$. Thus $\SPW(G)\geq |S|=2$.
%
``$\Rightarrow$'': 
Consider the graph $G$. Suppose that $\SPW(G)\geq k$, then there is a set $S=\{e_1,\ldots,e_k\}$ that is part of a minimal cut $C$ between $s$ and $t$. Denote $e_i=(a_i,b_i)$.
  
	By Prop.~\ref{prop:par_sets}, $G$ has a forward-subtree $T_s$ with root $s$ and leafs $\{a_i\}_{i\leq k}$, and a backward-subtree $T_t$ in $G$ with leaf $t$ and roots $\{b_i\}_{i\leq k}$.
	%Let $G_C$ be graph $G$ without the edges of $C$. Since $C$ is a minimal cut, there is a path from $s$ to every $a_i$, as otherwise we can remove the edge $e_i$ from $C$. The union of these paths is w.l.o.g. a forward tree, which we denote by $T_s$. We similarly build $T_t$ from the union of paths $b_i-t$. 
	%
%By Lemma~\ref{lemma:parallel_components},   there are disjoint components $T_s$ and $T_t$ that contains $\{s,a_1,\ldots,a_k\}$ and  $\{b_1,\ldots,b_k,t\}$, respectively. The minimal such component $T_s$ is a forward-subtree, since it is a union of paths from $s$ to $a_i$. Similarly, $T_t$ is w.l.o.g. a backward-subtree. 
%for every $i\leq k$, there is some path $p_i$ from $s$ to $t$ such that $e_i\in p_i$, and  $e_j\notin p_i$ for all $j\neq i$. 
%If there are several such paths, then $p_i$ is the path among them that has maximum edge overlap with the path $p_{i-1}$. 
%
Also, by definition of the parallel width, there is a \emph{simple} $s-t$ path $p'$ containing $S$, w.l.o.g. in that lexicographic order. %If there are several such paths, $p'$ is the one maximizing edge overlap with the union of $\{p_i\}_{i=1}^k$. 

 %Note that $a_{i+1}\neq b_i$ for all $i$, as otherwise the minimal cut between $s-t$ is either between $s$ and $b_i=a_{i+1}$ (in which case $e_{i+1}$ is redundant) or between $b_i=a_{i+1}$ and $t$ (in which case $e_i$ is redundant). We also argue that $a_i \neq a_j, b_i\neq b_j$ for all $i\neq j$, as otherwise the (simple) path $p'$ is cyclic, which is a contradiction.  Also, $s\neq a_i$ for all $i>1$ as otherwise $p_1 \cup p'$ form a cycle $s-a_1-b_1-\cdots-a_i=s$. Similarly, $t\neq b_i$ for all $i<k$. \rmr{this is the only place we assume $G$ is acyclic}
%
%Considering the topological order of nodes, we define the subgraph $G_i$ (for $i=2,\ldots,k$) as the induced subgraph of $G$ on all nodes between (and including) $b_{i-1}$ and $a_i$. 
%\rmr{this has to be re-written. There is a tree from s to S and from S to t, that needs to be contracted}
We now describe a series of d-minor operations on $G$ that will result in a variant of $G_{SP(k)}$.
Delete all edges and vertices that are not part of  $p'$, $T_s$ or $T_t$. % for some $i\leq k$. 
This leaves us with a graph $G'$ that is a valid subgraph of $G$ and thus,
%has the following form: The union of $\{p_i\}$ can be written as the union of $S$, a forward-subtree $T_s$ whose source is $s$ and whose leafs are $\{a_i\}_{i=1}^k$ (unless $a_1=s$); and a backward-subtree $T_t$ whose target is $t$ and whose sources are $\{b_i\}_{i=1}^k$ (unless $b_k=t$). Since $G'$ is a valid subgraph of $G$
 by Lemma~\ref{lemma:subgraph} and Thm.~\ref{thm:minor_embed}, is also a d-minor of $G$.

$p'$ is composed of a sequence of subpaths between vertices $s,y_1,x_2,y_2\ldots,x_{k-1},y_{k-1},x_k,t$, where each $x_i$ is the first intersection of $[p'_{b_{i-1},a_i}]$ with $T_s$. Thus, $x_i$ is an ancestor of (or coincides with) $a_i$ in $T_s$. Similarly, $\{y_i\}_{i=1}^{k-1}$ are on the backward-subtree $T_t$, where $y_i$ is the last intersection of $[p'_{b_i,a_{i+1}}]$ and $T_t$. Denote by $A_i\subseteq \{a_2,\ldots,a_k\}$ all leafs of the subtree of $T_s$ rooted by $x_i$, and   by $B_i\subseteq \{b_1,\ldots,b_{k-1}\}$ all roots of the subtree of $T_t$ whose leaf is $y_i$.
  In particular,  $a_i\in A_i$, and $a_j\notin A_i$ for $j<i$, as otherwise there is a cycle $x_i-a_j-b_j-y_j-x_i$. Likewise, $b_i\in B_i$ and $b_j\notin B_i$ for $j>i$. \rmr{cycles do not matter much. they only add variants where there is an additional vertex $s\notin A$, and there is no edge $(s,a_1)$. }

%We denote by $T_{s,i}$ the subtree of $T_s$ rooted in $x_i$. 
Note that the indegree of all nodes in $T_s$ is $1$, except for $\{x_i\}_{i=2}^k$ whose indegree is 2 (one edge from the parent in $T_s$, and one from the predecessor node on $p'$), and $s$ whose indegree is $0$. We thus backward-contract all edges in $T_s$ that do not point to some $x_i$. 
We get a forward-subtree $\hat T_s$:
\begin{itemize}[topsep=0.5ex,itemsep=0.1ex,labelindent=1em]
	\item The root of $\hat T_s$ is $s=x_1$, and its nodes are $\{x_i\}_{i=2}^k$;
	\item  Each path $x_i-a_i$ in $G'$ becomes a single node $x_i=a_i$ in $\hat G$;
	\item The subtree rooted by $x_i$ in $T_s$ becomes a subtree in $\hat T_s$ over nodes $A_i$ maintaining their order, i.e., children have higher index than their parent. For example, in Fig.~\ref{fig:SPk_var} on the right, $A_4=\{a_4,a_5\}$ and $a_4$ is a parent of $a_5$.   
	%\item The contracted graph contains an edge $(x_i,b_i)$ for all $i=1\ldots,k$ (these edges are not part of $\hat T_s$).
\end{itemize}
  We similarly contract $T_t$ to $\hat T_t$ on nodes $\{y_i\}$.
$\{y_i\}$ are the only nodes in $T_t$ whose outdegree is $>1$. After forward-contracting all edges of $T_t$ not originating in some $y_i$, we get a backward-subtree $\hat T_t$ over nodes $\{y_i\}_{i=1}^{k-1}$ whose leaf is $t=y_k$. Each subtree whose leaf is $y_i$ is contracted to a tree over nodes $B_i$ maintaining their order.

The last step is to contract every subpath $[p'_{y_i,x_{i+1}}]$ to a single edge $(y_i,x_{i+1})$. Denote the union of these edges by $\hat F$, so that $S\cup \hat F$ is the path we got after contracting $p'$. 

 We get that the contracted graph $\hat G' = S \cup \hat F \cup \hat T_s \cup \hat T_t$ is isomorphic to a variant of $G_{SP(k)}$. More specifically, $s$ and $t$ are isomorphic to themselves, each $x_i$ for $i=2,\ldots,k$ in $\hat G$ is isomorphic to $a_i$ in $G_{SP(k)}$, and each $y_i$ for $i=1,\ldots,k-1$ in $\hat G'$ is isomorphic to $b_i$ in $G_{SP(k)}$. For each $i=2,\ldots,k$, let $j$ be the maximal index such that $x_j$ is an ancestor of $x_i$ in $T_s$. If such $j$ exists, then the parent of $a_i$ in $\hat G'$ is $a_j$, and otherwise its parent is $s=a_1$. 
The parent of $a_i$ in $G_{SP(k)}$ is the closest ancestor $x_j$ of the node $x_i$ in $T_s$ (and similarly for the child of $b_i$).

%
%
%, and each $y_i$ is on the path $p_i$ between $b_i$ (included) and $t$. It is not possible that $x_i=s$ as this would mean that the simple path $p'$ contains a cycle $s-y_{i-1}-x_i$. Similarly, it is not possible that $y_i=t$. 
%
%
%It is sufficient to show that $G_{SP(k)}$ is d-embedded in $G'$.
%%For this, we argue that $G'$ is of one of the following forms. 
%We start from $G_{SP(k)}$ and specify a list of embedding operations that results in $G'$. 
%\begin{enumerate}
	%\item For $i\in[1,k-1]$ we subdivide the edge $(a_i,b_i)$ (and rename vertices) until we obtain the path $(a_i,b_i,\ldots,y_i)$; Note that the edge $(b_i,a_{i+1})$ now becomes $(y_i,a_{i+1})$.
	%\item For $i\in[2,k]$ we subdivide the edge $(a_i,b_i)$ until we obtain the path $(x_i,\ldots,a_i,b_i)$; Note that the edge $(y_{i-1},a_{i})$ now becomes $(y_{i-1},x_{i})$.
	%\item For $i\in[1,k-1]$ we subdivide the edge $(y_i,x_{i+1})$ until we obtain the path $y_i-x_{i+1}$.
	%\item For $i\in[2,k]$ we subdivide the edge $(s,x_i)$ until we obtain the path $s-x_i$.
	%\item For $i\in[1,k-1]$ we subdivide the edge $(y_i,t)$ until we obtain the path $y_i-t$.
	%\item If $s\neq a_1$ in $G'$, we subdivide the edge $(s,b_1)$ until we obtain the path $(s,\ldots,a_1,b_1)$.
	%\item If $t\neq b_k$ in $G'$, we subdivide the edge $(a_k,t)$ until we obtain the path $(a_k,b_k,\ldots,t)$.
%\end{enumerate}
%Each path $p_i$ in $G'$ is the result of steps 1,2,4,5 (also step~6 for $p_1$ and step~7 for $p_k$). The path $p'$ in $G'$ is the result of steps 1,2,3,6,7.

%``$2 \Rightarrow 3$'': Follows directly from Lemma~\ref{lemma:minor_embed}.

``$\Leftarrow$'': Follows directly from Lemma~\ref{lemma:embed} and Thm.~\ref{thm:minor_embed}, since $SPW$ for any variant of $G_{SP(k)}$ is $k$.
\end{proof}

Since $G_{SP(k)}$ has $2k$ vertices, we get the
following bound:
\begin{corollary}
For any TDAG $G=\tup{V,E,s,t}$, $\SPW(G)\leq \frac{|V|}{2}$. 
\end{corollary}
Another corollary of Theorem~\ref{th:SPW_embed} is a generalization of the lower bounds on negative externality from \cite{babaioff2007congestion,MP18}. These papers show how  instances with high externality (depends on $k$)  can be constructed on any variant of $G_{SP(k)}$.  By Theorem~\ref{th:SPW_embed} this is true for \emph{any graph} $G$ with $SPW(G)\geq k$.

\subsection{Series-parallel graphs}
Series-parallel 2-terminal graphs have been long studied in contexts such as electric circuits~\cite{Duffin65}, complexity of graph algorithms~\cite{takamizawa1982linear}, and also routing games~\cite{milchtaich2006network,epstein2009efficient}. %Whereas most papers assume undirected graphs, there is a very similar definition for directed graphs due to \cite{jakoby2006space}, thus we bring them together.
\begin{definition}[Series-parallel graph~\cite{eppstein1992parallel,jakoby2006space}]\label{def:DSP}
A \emph{[directed] series-parallel graph}  is a 2-terminal graph $\tup{V,E,s,t}$, and is either a single edge $(s,t)$, or is composed recursively by one of the two following steps:
\begin{description}[topsep=0.5ex,itemsep=0.1ex,labelindent=1em]
\item[Serial composition.] Combine two [directed] 2-terminal graphs $\tup{V_1,E_1,s_1,t_1},\tup{V_2,E_2,s_2,t_2}$ serially by merging $t_1$ with $s_2$.
\item[Parallel composition.] Combine two [directed] 2-terminal graphs $\tup{V_1,E_1,s_1,t_1},\tup{V_2,E_2,s_2,t_2}$ in parallel by merging $s_1$ with $s_2$, and $t_1$ with $t_2$.
\end{description}
\end{definition}

Our last result in this section is showing that directed series-parallel graphs (DSP) characterize exactly the 2-terminal graphs with serial-parallel width of 1. % While we could show this directly, it will be more insightful to 
%We prove this using  known characterizations of directed and undirected series-parallel graphs, thereby also shed some light on the connections between embeddings, d-embeddings, and subdivisions.
 %\paragraph{Directed graph subdivisions}

\begin{proposition}[\cite{holzman2003network}]\label{prop:holzman} Let $G$ be a 2-terminal directed graph. Then $G$ is a DSP if and only if $G_B$ is not h-embedded in $G$. % has no subgraph that is isomorphic to a subdivision of the Braess graph $G_B$. 
\end{proposition}

Proposition~\ref{prop:holzman} and the relation between h-embeddings and d-embeddings, gives the following. 
\begin{theorem}\label{TH:DSP_EMBED}
Let $G$ be a TDAG, and let $k\geq 2$. The following conditions coincide.
%(1) $G$ is a directed series-parallel graph.
%(2) The directed Braess graph $G_B$ is not d-embedded in $G$.
%(3) $\SPW(G)= 1$.
\begin{enumerate}[topsep=0.5ex,itemsep=0.1ex,labelindent=1em]
  \item $G$ is a directed series-parallel graph.
	\item The directed Braess graph $G_B$ is not d-embedded in $G$.
	\item $\SPW(G)= 1$.
\end{enumerate}
%If a 2-terminal directed graph $G$ is a DSP, then Braess graph $G_B$ is not d-embedded in $G$.
\end{theorem}
\begin{proof}
Note that $G_B$ has no hubs, as all vertices have at most 3 neighbors. Thus by Prop.~\ref{prop:embed_finite}, $G_B$ is d-embedded in $G$ if and only if it is h-embedded (as $|J|=0$, $\calG$ contains only $G_B$ itself). Thus we get (1)$\iff$(2).  

 (2)$\!\iff\!$(3) follows as a special case from Thm.~\ref{th:SPW_embed}.
\end{proof}

\adl{there is a whole hidden section here. I think it is redundant.}
\noijcai{

\paragraph{Undirected graph embeddings}

\begin{proposition}[\cite{Duffin65,milchtaich2006network}] \label{prop:milch}
For an undirected 2-terminal graph $H$, the following conditions coincide: 1. $H$ is series-parallel; 2. for every pair of distinct vertices $u$ and $v$, if $u$ precedes $v$ in some path $p$ containing both
vertices, then $u$ precedes $v$ in all such paths; and 3. the (undirected) Braess graph\footnote{This graph is also known as the \emph{Wheatstone network}~\cite{milchtaich2006network}.} $\ol G_B$ is not h-embedded in $H$.  
\end{proposition}
We can rephrase condition~(2) above, to get two more equivalent conditions that will be useful. 
\begin{lemma}\label{lemma:milch2}
For an undirected 2-terminal graph $H$, the following conditions coincide: 1. $H$ is series-parallel; 4.
there are no two paths in $H$ that go through the same edge in opposite directions; and 5. there is a unique 2-terminal directed graph $G$ such that $\ol G=H$.
\end{lemma}
\begin{proof}
We use the equivalence of conditions (1) and (2) from Proposition~\ref{prop:milch}. 
That condition (2) entails (4) is obvious, as the existence of such a bidirectional edge $(a,b)$ entails the existence of a pair $\{u=a,v=b\}$. 
(4) entails (5): Suppose there are two directed 2-terminal graphs $G,G'$ such that $\ol G=\ol G'=H$. This means there is some edge $(a,b)\in E$ such that $(b,a)\in E'$. Since the graphs are 2-terminal, there is a directed $s-t$ path $p$ in $G$ that contains $(a,b)$, and likewise a directed $s-t$ path $q$ in $G'$ that contains $(b,a)$. Both of $p,q$ are valid $s-t$ paths in $H$ that go through $(a,b)$ in opposite directions. 

Finally, (5) entails (2), since if there is a unique way to direct the edges of $H$, then every $s-t$ path in $H$ uniquely determines an ordering of all vertices along the path.  
\end{proof} 
Unfortunately, since d-embedding allows for a different set of operations than h-embedding (as per Def.~\ref{def:h_embed}), we cannot use Prop.~\ref{prop:milch} directly. However we can consider a modified definition as follows.
\footnote{A reversed definition using edge removal and edge contraction appears in the working paper version of \cite{epstein2009efficient}. The published version used Def.~\ref{def:h_embed} .}
%define an alternative notion of undirected graph embedding that is more similar to d-embedding.
%We could think of an undirected version of the above definition, where an arbitrary split of nodes is allowed.  
\begin{definition}\label{def:s_embed}
A 2-terminal undirected  graph $H'$ is \emph{split-embedded} (or s-embedded) in $H$ if $H'$ is isomorphic to a graph
derived from $H$ by applying the following operations any number of times in any order:
\begin{description}[topsep=0.5ex,itemsep=0.1ex,labelindent=1em]
\item[Addition.] Add a new edge joining two existing vertices;
	\item[Terminal extension.]  Add a new edge $e$ joining $s$ or $t$ with another, new vertex, which becomes the new source or target, respectively.
	\item[Split.] Replace a non-terminal node $a$ by two nodes $a'$ and $b'$ and an edge $(a',b')$, such that each new node retains at least one edge of $a$.
\end{description}
\end{definition}

\begin{lemma}\label{lemma:s_embed_closed}
The set of 2-terminal undirected graphs is closed under s-embedding operations.
\end{lemma}

Interestingly, the characterization of series-parallel networks in \cite{milchtaich2006network} still holds with this more lax definition of s-embedding, without changing Milchtaich's proof. From that proof and Lemma~\ref{lemma:milch2} we get the following. For completeness, the full proof is in the appendix.  

\begin{proposition}\label{prop:milch2}
For an undirected 2-terminal graph $H$, the following conditions coincide: 
\begin{enumerate}[topsep=0.5ex,itemsep=0.1ex,labelindent=1em]
	\item $H$ is series-parallel;
	\item for every pair of distinct vertices $u$ and $v$, if $u$ precedes $v$ in some path $p$ containing both
vertices, then $u$ precedes $v$ in all such paths;
	\item  the (undirected) Braess graph $\ol G_B$ is not \textbf{s-embedded} in $H$; 
	\item there are no two paths in $H$ that go through the same edge in opposite directions; 
	\item there is a unique 2-terminal directed graph $G$ such that $\ol G=H$.
\end{enumerate}
\end{proposition}

%For a directed graph $G$, we denote by $\ol G$ the undirected variant of the same graph. 
There is a natural correspondence between d-embeddings and s-embeddings. This is since  we can replicate the sequence of d-embedding operations on $G'$ with s-embedding operations on $\ol G'$. %The other direction also holds in some sense:

\begin{observation}\label{ob:embed_undirected}
If $G'$ is d-embedded in $G$ then $\ol G'$ is s-embedded in $\ol G$.
\end{observation}
This is obvious since for every d-embedding step on $G'$ we can apply a corresponding s-embedding step on $\ol G'$. However, the other direction does not hold in general. Fig.~\ref{fig:no_d_embed} shows a counter example. 

\begin{figure}
\input{no_d_embed}
\caption{\label{fig:no_d_embed}
The graph $G'$ is not d-embedded in $G$, since $x$ retains an outgoing edge and $y$ retains an incoming edge (indeed, $G'$ has a path $a-b$ that does no exist in $G$). In contrast, splitting $x$ to the edge $(x,y)$ is a valid s-embedding step in $\ol G'$.}
\end{figure}

\begin{lemma}\label{lemma:DSP}
$G$ is a DSP if and only if $\ol G$ is series-parallel.
\end{lemma}
\begin{proof}
Assume by induction that this is true for any graph with $k$ edges (the base case of $k=1$ is obvious). If $G$ is a DSP then  it is constructed of two smaller DSPs $G_1,G_2$ in series [in parallel]. By induction, $\ol G_1,\ol G_2$ are series-parallel. We can construct $\ol G$ by joining $\ol G_1,\ol G_2$ in series [resp., in parallel]. 

In the other direction, suppose that $H=\ol G$ is series-parallel.  Consider the last (series/parallel) construction step of $H$ from $H_1,H_2$. Direct the edges of $H_1,H_2$ to obtain directed graphs $G_1,G_2$. By induction,  $G_1,G_2$ are DSPs, and thus joining them (in series/parallel) results in a DSP $G_{1+2}$ such that $\ol G_{1+2}=H=\ol G$. Finally, by conditions (1) and (5) of Lemma~\ref{lemma:milch2} there is only one way to direct the edges of $H$ from $s$ to $t$ and thus $G=G_{1+2}$.
\end{proof}

%The (directed) Braess graph $G_B$ is the 4-vertices graph in Fig.~\ref{sfig:Braess} (ignore the costs).
We can now turn to prove a  characterization of DSPs in terms of their parallel-width. %\footnote{In \cite{holzman2003network} there is a similar characterization of \emph{extension-parallel graphs}, as graphs where there are no two edges $e,e'$ and three paths such that one uses only $e$, one uses only $e'$ and the third uses both. Note that this condition implies a parallel-width of $1$ but not vice versa, as such $e,e'$ may not be contained in any minimal cut.}
It will follow as an easy corollary of the previous results in the paper. 
\begin{theorem}\label{TH:DSP_EMBED}
Let $G$ be a TDAG, and let $k\geq 2$. The following conditions coincide:
\begin{enumerate}[topsep=0.5ex,itemsep=0.1ex,labelindent=1em]
  \item $G$ is a directed series-parallel graph;
	\item The directed Braess graph $G_B$ is not d-embedded in $G$;
	\item $\SPW(G)= 1$;
	\end{enumerate}
%If a 2-terminal directed graph $G$ is a DSP, then Braess graph $G_B$ is not d-embedded in $G$.
\end{theorem}
\begin{proof}
%, thus this is an ``if and only if'' entailment:
%\labeq{DSP}
%{\text{}.}
``$1\Rightarrow 2$'':
 Suppose that $G_B$ is d-embedded in $G$. Then by Observation~\ref{ob:embed_undirected}, $\ol G_B$ is s-embedded in $\ol G$. Due to Prop.~\ref{prop:milch2} (conditions (1) and (3)), this means that $\ol G$ is not series-parallel. However this means that $G$ is not a DSP by Lemma~\ref{lemma:DSP}. 

``$2 \Rightarrow 1$'': Suppose that $G$ is not a DSP. Then by Prop.~\ref{prop:holzman}, $G_B$ is h-embedded in $G$. % is isomorphic to a subdivision of a supergraph of the Braess graph  $G_B$.
 Then, by Lemma~\ref{lemma:subdivide}, $G_B$ is d-embedded in  $G$. 

``$2\iff 3$'': Follows directly from Thm.~\ref{th:SPW_embed}.
\end{proof}
%
%By Lemma~\ref{??}, we get the last corollary:
%\begin{corollary}
%A 2-terminal directed graph $G$ is a DSP if and only if $\SPW(G)=1$.
%\end{corollary}

%
%\begin{figure}
%\input{network_fig_PCk_undirected}
%\caption{Consider the undirected version $\ol G$ of the graph on the right. It is a valid network since any edge belongs to an $s-t$ path (though not to a simple path!). It is clearly not series-parallel, and thus by Milchtaich, the Braess graph $\ol G_B$ is embedded in $\ol G$. }
%\end{figure}

It can be similarly shown that a TDAG is \emph{extension parallel} if and only if it excludes the Braess graph and the graph $\{(s,a),(s,a),(a,t),(a,t)\}$, which in turn holds if and only if there is no serial-concurrent set larger than 1.
\rmr{add proof}
}
%%%%%%++++++++++++++
%%%%%%++++++++++++++

\section{Computational Problems}\label{sec:comp}

\newcommand{\kvdp}{$m$-\textsc{VertexDisjointPaths}\xspace}
\newcommand{\vdp}{\textsc{VertexDisjointPaths}\xspace}
\newcommand{\edp}{\textsc{EdgeDisjointPaths}\xspace}
\newcommand{\isMinor}{\textsc{IsMinor}\xspace}
\newcommand{\isDMinor}{\textsc{IsDMinor}\xspace}
\newcommand{\isDEmbedded}{\textsc{IsDEmbedded}\xspace}
We first ask whether we can efficiently decide when a directed graph is 2-terminal.
\begin{proposition}
\label{prop:tdag-dec}
It is \NP-complete to decide if a directed graph is 2-terminal, but in \PP if the graph is acyclic.
\end{proposition}
If a graph is acyclic, from Lemma~\ref{lem:tdag-char} we get that it is 2-terminal, i.e. it is a TDAG, if it has a unique source and a unique sink. Clearly, we can check this in polynomial time. The \NP-completeness result follows  from the combination of Lemma~\ref{lem:twotermhard} with 
Observation~\ref{obs:two-term}.

\begin{lemma}
\label{lem:twotermhard}
Given a directed graph it is \NP-complete to decide if there exists a simple $s-t$ path that 
goes through an edge $(u,v)$.
\end{lemma}
\begin{proof}
The containment is obvious.
The completeness follows from the \tpath problem~\cite{ForHopWyl80}. An instance of \tpath is consisted by a directed graph $G$ and four
distinct vertices $u_1, u_2, v_1, v_2$ and we are asked if $G$ has disjoint paths 
connecting $u_1$ to $v_1$ and $u_2$ to $v_2$. So, if we add the edge $(v_1,u_2)$ we
get that it is \NP-complete to decide if there exists a simple $s-t$ path that goes through
edge $(v_1,u_2)$. 
\end{proof}

\begin{observation}
\label{obs:two-term}
An edge $(u,v)$ belongs to a simple $s-t$ path if and only if there exist simple $s-u$ and 
$v-t$ paths that are vertex disjoint.
\end{observation}
\begin{proof}
If $u-v$ belongs to a simple $s-t$ path, then such a path can be written as 
$[p_{su}]-[p_{vt}]$ where the paths $[p_{su}]$ and  $[p_{vt}]$ are vertex disjoint, since
the $s-t$ path is simple.
For the other side, if there exist vertex disjoint paths $[p_{su}]$ and  $[p_{vt}]$,
then clearly the path $[p_{su}]-[p_{vt}]$ is simple.
\end{proof}

The next two natural computational questions accept as input 2-terminal graphs $G$ and $G'$.
\begin{description}
	\item[\isDMinor]: is $G'$ a d-minor of $G$?
	\item[\isDEmbedded]: is $G'$ d-embedded in $G$?
\end{description}
The complexity may depend on whether the graphs are TDAGs (in which case the questions coincide), and also on whether $G'$ is a fixed graph of size $k$. We write down some of our results explicitly, and summarize all of them in Table~\ref{tab:is}.
% Our results for \isDMinor and all other questions are summarized in Tables~\ref{tab:is}, \ref{tab:max}.

%\vspace{-3mm}
\subsection{Testing properties of edge sets}
%\paragraph{\bf Testing properties of edge sets.}
We are interested in the following questions on a given 2-terminal graph $G=\tup{V,E,s,t}$ and a set $S=\{(a_i,b_i)\}_{i\leq k}$ of  $k$ edges:
\begin{description}
	\item[\iss]: is there an $s-t$ path containing $S$?
	\item[\isc]: is $S$ concurrent?
	\item[\isp]: is $S$ parallel?
    \item[\issc]: is $S$ both serial and concurrent?
	\item[\issp]: is $S$ both serial and parallel?
\end{description}

\begin{table*}
\centering
\scalebox{0.9}{
\begin{tabular}{|l|c|c|c|c|}
\hline
& \multicolumn{2}{c|}{2-terminal graph} &\multicolumn{2}{c|}{TDAG}\\
        &  any $k$                         & fixed $k$  & any $k$ & fixed $k$ \\
				\hline
\iss    & \NP-c   &  \NP-c [P.~\ref{prop:iss_hard}] &  \PP  \noijcai{[P.\ref{prop:iss_TDAG_p}]} & \PP \\
\isc    & ?       &  ?                            &  \PP  [P.\ref{prop:iss_TDAG_p}] & \PP \\
\isp    &  ?      & ?                             &  ?                              & \PP [P.~\ref{prop:isp_TDAG_k_p}] \\
\issc    & \NP-c & \NP-c [P.~\ref{prop:iss_hard}] &  \PP [P.\ref{prop:iss_TDAG_p}]  & \PP \\
\issp    & \NP-c & \NP-c [P.~\ref{prop:iss_hard}] &  ?                              & \PP [P.~\ref{prop:isp_TDAG_k_p}] \\
\hline
\maxs    & \NP-c   [P.~\ref{prop:maxs_hard}] &  \PP [P.~\ref{prop:maxs_k_p}] &  \PP [P.~\ref{prop:maxs_TDAG_p}] & \PP \\
\maxc    & ?   &  ?    & ? & \PP [check all]\\
\maxp    &  \NP-c & ? &  \NP-c [P.~\ref{prop:maxp_TDAG_hard}]  & \PP [C.~\ref{cor:minors}] \\
\maxsc    & \NP-c [P.~\ref{prop:maxsc_np}] & ? &  ? & \PP \\
\maxsp    & ? & ? &  ? & \PP [C.~\ref{cor:minors}] \\
\hline
\isDMinor  & \NP-c & ? & \NP-c [P.\ref{prop:isminor_TDAG_hard}] & \PP [T.~\ref{thm:isembedded}]\\
\isDEmbedded & \NP-c &   \PP * &                 \NP-c [P.\ref{prop:isminor_TDAG_hard}] & \PP [T.~\ref{thm:isembedded}] \\
\hline
\end{tabular}}
\caption{\label{tab:is}The computational complexity of problems we study. \newline
 * - \isDEmbedded in easy if the minor $G'$ is acyclic.}
%Computational problems for a subset of edges $S$ of size $k$.}
\end{table*}

Note that since all of these properties are phrased in terms of existence, containment in \NP is trivial.  

Our main tool in many of the results, both positive and negative, will be the \kvdp problem: given a directed graph $G=\tup{V,E}$ and $m$ pairs of vertices $\{(x_i,y_i)\}_{i\leq m}$, find whether there are vertex-disjoint paths $x_i-y_i$ in $G$ for all $i\leq m$. This problem is equivalent to that of checking if a graph $G'$ is h-embedded in $G$~\cite{ForHopWyl80}, yet using it for our problems requires some modifications. 
%The problem is known to be 
The problem is \NP-complete even when $G$ is a DAG~\cite{vygen1995np},
%\footnote{The result in the paper is for edge-disjoint paths, but there is a reduction from $2$-\edp to $2$-\vdp on a DAG~\cite{tholey2006solving}.} 
and \NP-complete for $m=2$ in general directed graphs~\cite{ForHopWyl80}. In contrast, it is in \PP when $G$ is a DAG \emph{and} $m$ is fixed~\cite{ForHopWyl80}.  
The runtime of the algorithm however, is exponential in $m$, in contrast to the famous $O(|V|^3)$ algorithm for \kvdp with fixed $m$ (and \isMinor of fixed minor $H'$)  on \emph{undirected} graphs~\cite{robertson1995graph}. At least for the $m$-\edp problem on DAGs it was shown that such a fixed-parameter tractable algorithm does not exist under standard assumptions~\cite{slivkins2010parameterized}. Recently, there was some progress on the \kvdp problem with fixed $m$ for general directed graphs~\cite{kawarabayashi2015towards}.

\begin{proposition}\label{prop:iss_hard}
\iss, \issc and \issp are \NP-complete even for $k=3$.
\end{proposition}
For $k=1$ every instance is a `yes' instance, as any single edge is part of a simple path and part of a minimal cut.

\begin{proof}
For hardness, we use a reduction from \kvdp with $m=2$. Consider an instance $G$ and pairs $(x_1,y_1), (x_2,y_2)$ of \kvdp.
We define a 2-terminal graph $G'$ as follows. We add two new nodes $s$ and $t$ to $G$, and add edges from $s$ to all of $V$,  from all of $V$ to $t$, and the edge $(y_1,x_2)$.  

We define the set $S=\{(s,x_1),(y_1,x_2),(y_2,t)\}$. Note that $S$ is a parallel set since it is a minimal cut in the subgraph restricted to nodes $\{s,x_1,y_1,x_2,y_2,t\}$, which is a valid subgraph of $G'$. By Lemma~\ref{lemma:concurrent}, $S$ is also concurrent.

Suppose there are disjoint  paths $p_1=x_1-y_1$ and $p_2=x_2-y_2$, then there is a simple $s-t$ path $s-[p_1]-[p_2]-t$ that contains $S$, which means $S$ is serial (and thus also serial-parallel and serial-concurrent). In the other direction, if $S$ is serial, there is a simple path containing all edges of $S$. Edges must appear in that order where $(s,x_1)$ is first and $(y_2,t)$ is last, which means there must be vertex disjoint paths $x_1-y_1$ and $x_2-y_2$.
\end{proof}

\begin{proposition}\label{prop:iss_TDAG_p}
\iss, \isc, \issc on a TDAG are in \PP.
\end{proposition}
\begin{proof}
We first prove for \iss. 
Let $S=\{(a_i,b_i)\}_{i \leq k}$ be the set we want to check. We sort the vertices of $G$ with a topological sort, and sort $S$ according to the order of the $a_i$ vertices in the sorted graph. We then check if there is a path from $s$ to $a_1$, from $b_1$ to $a_2$, and so on. If such paths exist, they must be disjoint since they are using only forward edges in the sorted graph. Similarly, if $S$ is serial then it is contained in some $s-t$ path $p$. Edges of $S$ must appear in $p$ in the above order, as otherwise $p$ contains a backward edge which cannot exist in the sorted graph.

 For \isc, we need to check for every edge $e\in S$ that there is an $s-t$ path containing $e$ in the graph $G_e=G\setminus(S\setminus\{e\})$. This means checking if the set $\{e\}$ is serial in $G_e$ for every $e\in S$, which is trivial. The algorithm for \issc follows from the previous two. 
\end{proof}

The most tricky part is the complexity of identifying a parallel set. Using some of the structural results obtained in the previous sections, we can show the following.

	\begin{proposition}\label{prop:isp_TDAG_k_p} 
 \isp is in \PP for TDAGs and fixed $k$.	
%	\!\!\isp is in \!\PP\! for \!TDAGs and fixed $k$.
	\end{proposition}
	\noijcai{
	\begin{proof}[Proof sketch]
	%Denote $e_i=(a_i,b_i)$ for any $e_i\in S$. 
	Denote the vertices of $S$ by $A=\{a_1,\ldots,a_k\}$ and $B=\{b_1,\ldots,b_k\}$.
	By Prop.~\ref{prop:par_sets}, it suffices to decide if $G$ contains a forward-subtree $T_s$ to all of $A$, and a backward-subtree $T_t$ from all of $B$ to $t$. Our algorithm ``guesses'' the structure of $T_s$ and $T_t$, that is, which nodes are the ``junctions'' with more than one child or parent in its respective tree, and to which other junctions they are indirectly connected. Crucially, there is only a fixed number of such guesses for any fixed $k$. 
	Then, we duplicate each (conjectured) junction of $T_s$ according to its (conjectured) number of direct descendant junctions, and likewise for $T_t$ according to the number of direct ancestors. Finally, we use the \kvdp algorithm of \cite{ForHopWyl80} to find whether such paths among all conjectured junctions indeed exist. If such paths are found for some guess of junctions, we obtain $T_s$ and $T_t$ by merging back duplicates of each junction, and $T_s,T_t$ are guaranteed to be disjoint. In the other direction, clearly if such $T_s,T_t$ exist, then one of our guesses will find them.
	\end{proof}
\noijcai{	
	Specifically, let $G$ be a TDAG, and $k$ be a constant. We can decide in time polynomial in the size of $G$ if a set $S$ of $k$ edges is parallel (or serial-parallel). }
	\end{proposition}}
	
\begin{proof}

\if 0

\begin{algorithm}[htb]
\caption{IsParallel \label{alg:isp}} 
\begin{algorithmic}[1]
%% \REQUIRE{}
%% \ENSURE{}
%%  
%% \medskip
\STATE{INPUT: a TDAG $G=\tup{V,E,s,t}$, a set $S=\{(a_i,b_i)\}_{i\leq k}$ of edges.}
\STATE{Set $A=\{a_i\}_{i\leq k}, B=\{b_i\}_{i\leq k}$;} 
\FOR{all subsets $X,Y\subseteq V$ of size $k-1$}
    \STATE{Define graph $G_X=\tup{\{s\} \cup X \cup A,E_X}$, where there is an edge $(v,u)$ in $E_X$ if there is a $v-u$ path in $G$;}
		\STATE{Define graph $G_Y=\tup{\{t\} \cup Y \cup B,E_Y}$ similarly};
		\FOR{every forward-tree $T_X$ that spans $G_X$ and every backward-tree $T_Y$ that spans $G_Y$}
				\STATE{Let $d(x_j)$ be the outdegree of every $x_j\in X$ in $T_X$;}
				\STATE{Let $d(y_j)$ be the indegree of every $y_j\in Y$ in $T_Y$;}
				\STATE{create a new set of nodes $\hat V$ from $V$ as follows:}
				\STATE{Replace each $x_j\in X$ with $d(x_j)+1$ nodes $x_j^0,x_j^1,\ldots,x_j^{d(x_j)}$;}
				\STATE{Replace each $y_j\in Y$ with $d(y_j)+1$ nodes $y_j^0,y_j^1,\ldots,y_j^{d(y_j)}$;}
				\STATE{Create a graph $\hat G=\tup{\hat V,\hat E}$ as follows:}
				\STATE{$\hat E$ contains all edges of $E$ not connected to $X,Y$;}
				\STATE{The incoming edges of $x_j^0$ in $\hat E$ are all incoming edges of $x_j$ in $E$;}
				\STATE{The outgoing edges of each $x_j^\ell$ in $\hat E$ are all outgoing edges of $x_j$ in $E$;}
				\STATE{The outgoing edges of $y_j^0$ in $\hat E$ are all outgoing edges of $y_j$ in $E$;}
				\STATE{The incoming edges of each $y_j^\ell$ in $\hat E$ are all incoming edges of $x_j$ in $E$;}
				\STATE{Define a set $\hat S$ of pairs, where each pair is an edge of $T_X$ or $T_Y$;}
				\STATE{Find vertex-disjoint paths between all pairs of $\hat S$ in $\hat G$;}
				\IF{found such a set of paths}
				   \RETURN{TRUE}
		\ENDFOR
	\ENDFOR
	\RETURN{FALSE}
\end{algorithmic}
\end{algorithm}
Consider the  Alg.~\ref{alg:isp}. 

\else
	Denote $e_i=(a_i,b_i)$ for any $e_i\in S$.  Denote $A=\{a_1,\ldots,a_k\}$ and $B=\{b_1,\ldots,b_k\}$.
	By Prop.~\ref{prop:par_sets}, it suffices to decide if $G$ contains a forward-subtree $T_s$ to all of $A$, and a backward-subtree $T_t$ from all of $B$ to $t$.   Note that $T_s$ contains at most $k-1$ ``junctions'', i.e., nodes with outdegree greater than one (including $s$). Suppose first that we guess what these vertices are and what is their hierarchy, and denote them by $X=\{s=x_1,\ldots,x_{k'}\}$ and relations $T_X$. We similarly guess a set $Y$ of junctions in $T_t$ and the relations among them $T_Y$.  Our algorithm works as follows: 
	
	\begin{itemize}
		\item For every $x_j$ with degree  $d_j$ in $T_X$, split $x_j$ into $d_j+1$ nodes such that one of them $x^{0}_j$ retains all incoming edges (entry port), and each of the other $x^{v_j}_j$ (exit port) retains all outgoing edges. $v_j$ is the first node from $X\cup A$ downward from $x_j$ on $T_s$.
		\item  Connect $x^{0}_j$ to all of $x^{v_j}_j$.
		\item Similarly split each $y_j\in Y$ to multiple entry ports and a single exit port.
		\item Find vertex-disjoint paths from each exit port to the entry port of one child in $T_X$ or $T_Y$, respectively. E.g. from 
		$x^{v_j}_j$ to  $a_i$ if $v_j=a_i$ for some $i\leq k$, or to $x^0_{j'}$ if $v_j=x_{j'}$ for some $j'\leq k'$. 
	\end{itemize}
%	Consider the algorithm above. 

	\fi
	
	The total number of edges in each tree $T_X,T_Y$ is at most $2k$,  so the total number of paths we seek in each iteration is less than $4k$. Such paths, if exist, can be found in time $|V|^{O(k^2)}$ due to the result of~\cite{ForHopWyl80}. 
	
	If such vertex-disjoint paths exist, then merging back all copies of each junction will provide us with a disjoint forward-subtree $T_s$ and backward-tree $T_t$. In the other direction, if such trees exist and use junctions $X$ and $Y$ respectively, then the paths between every two junctions are vertex-disjoint except in the junctions themselves. Since we split each junction, these paths will be fully vertex disjoint. Thus the algorithm will always find trees $T_s,T_t$ using junctions $X,Y$, if such exist. 
	
	The total number of iterations is the number of ways to select $2k$ vertices out of $|V|$, times the number of trees we can try on each set of size $2k$ (less than $(2k)^{(2k)}$ by Cayley's formula), so in total no more than $|V|^{O(k^2)}$ iterations.
	
%	The total runtime is $|V|^{O(k^2)}$ which is polynomial for fixed $k$. 
	%To find if such trees exist at all, we only need to run the algorithm for every possible guess of $X$ and $Y$. This means selecting less than $2k$ junctions out of $n$ vertices, which is polynomial for a fixed $k$. 
	
	\medskip Since by Prop.~\ref{prop:iss_TDAG_p} we know how to determine if $S$ is serial in polynomial time (even polynomial in $k$ for TDAGs), we can check whether it is serial-parallel by checking each property separately. 
	\end{proof}

\subsection{Testing width properties of graphs}
%\vspace{-3mm}
%\paragraph{\bf Testing width properties of graphs.}
%We are interested in the following questions on a 
Given 2-terminal graph $G$ and an integer $k$ we study  the following questions .
%All problems are in \NP although containment in \NP is not always completely trivial. We use the fact that whenever the problem \textsc{IsZZZ}\xspace is in \PP for a fixed $k$, so is \textsc{MaxZZZ}\xspace, by checking every subset $S$ of size $k$. 
\begin{description}
	\item[\maxs]: is there a serial set $S$ of size $\geq k$?
	\item[\maxc]: is there a concurrent set $S$ of size $\geq k$?
	\item[\maxp]: is there a parallel set $S$ of size $\geq k$? %(equivalently, is $PW(G)\geq k$?
   \item[\maxsc]: is there a set $S$ of size $\geq k$ that is both serial and concurrent?
	\item[\maxsp]: is there a set $S$ of size $\geq k$ that is both serial and parallel? %(equivalently, is $SPW(G)\geq k$?
\end{description}
\noijcai{
\begin{table}
\centering
\scalebox{0.7}{
\begin{tabular}{|l|c|c|c|c|}
\hline
& \multicolumn{2}{c|}{2-terminal graph} &\multicolumn{2}{c|}{TDAG}\\
        &  any $k$                         & fixed $k$  & any $k$ & fixed $k$ \\
				\hline
\maxs    & \NP-c   [P.~\ref{prop:maxs_hard}] &  \PP [P.~\ref{prop:maxs_k_p}] &  \PP [P.~\ref{prop:maxs_TDAG_p}]  & \PP \\
\noijcai{ \maxc    & ?   &  ?    & ? & \PP [check all]\\}
\maxp    &  \NP-c & ? &  \NP-c [P.~\ref{prop:maxp_TDAG_hard}]  & \PP [C.~\ref{cor:minors}] \\
\noijcai{ \maxsc    & \NP-c [P.~\ref{prop:maxsc_np}] & ? &  ? & \PP \\}
\maxsp    & ? & ? &  ? & \PP [C.~\ref{cor:minors}] \\
\hline
\isDMinor  & \NP-c & ? & \NP-c [P.\ref{prop:isminor_TDAG_hard}] & \PP [T.~\ref{thm:isembedded}]\\
\isDEmbedded & \NP-c &   \PP *  &                 \NP-c [P.\ref{prop:isminor_TDAG_hard}] & \PP [T.~\ref{thm:isembedded}] \\
\hline
\end{tabular}}
\caption{\label{tab:max}Does graph $G$ has a subset of edges $S$ of size at least $k$ with a given property. * - \isDEmbedded in easy if the minor $G'$ is acyclic.}
\end{table}
}

\begin{proposition}\label{prop:maxs_hard}
\maxs is \NP-complete.
\end{proposition}
\begin{proof}
Containment in \NP is easy since a witness is a set $S$ of size $k$ and an $s-t$ path containing $S$. 
For hardness, we use a reduction from \iss for $k=3$. Given a 2-terminal graph $G$ and set $S=\{e_1,e_2,e_3\}$, we subdivide each edge in $S$ $4|V|$ times and set $k=10|V|$. The new graph has a path of length $k$ (and thus a serial set of size $k$) if and only if this path use all three subdivided edges.  
\end{proof}

Interestingly, \maxs is easier than \iss for fixed $k$. 
\begin{proposition}\label{prop:maxs_k_p} 
\maxs is in \PP for fixed $k$. %can be solved in $O(|V|^k)$ time.
\end{proposition}

\begin{proof}
%Note that there is a serial set of size $k$ if and only if there is an $s-t$ path of length at least $k$. Such a path can be written as $p=s-x-t$, where $|p_{sx}|=k$, and $p_{xt}$ does not vertex-intersect $p_{sx}$. We thus use the following algorithm.
We begin from the source $s$, and start listing all simple paths of length 1, length 2 and so on until length $k$. The number of such paths is a most $n^k$. If we found no path of length $k$, then we are done and this a ``no'' instance. Otherwise, for each path $q=s-x$ of length $k$, we remove the internal vertices of $q$ from $G$ and search for an arbitrary simple path $q'=x-t$, which can be done via Dijkstra. If such a path exists then $[q]-q'-t$ is a simple $s-t$ path, and $q$ is a serial set of size $k$. 

In the other direction, if a serial set $S$ of size $k$ exists, it is contained in some $s-t$ path $p=(s,x_1,\ldots,x_k,\ldots,t)$. Our algorithm is guaranteed to list down $q=[p_{s,x_k}]$, and then find a path $x_k-t$ in $G\setminus \{q\}$.  
\end{proof}

\begin{proposition}
\label{prop:maxs_TDAG_p}
%\maxs is in \PP on a TDAG. A serial set $S$ can be found in time $O(|V|^2)$.
\maxs on a TDAG can be solved in time $O(|V|^2)$.
\end{proposition}

\begin{proof}
The problem is equivalent to finding the longest $s-t$ path in $G$. This is known to be possible by setting the weight of all edges to $-1$ and use a weighted shortest path algorithm such a Dijkstra. The condition for such an algorithm to work is that there are no negative cycles, which trivially holds on a TDAG. 
\end{proof}

\begin{proposition}
\label{prop:maxp_TDAG_hard}
\maxp is \NP-complete even on TDAGs.
%\lcut problem is \NP-complete even on 2-terminal directed acyclic graphs.
\end{proposition}
\noijcai{
\begin{proof}[Proof sketch]
We reduce from Max Directed Cut on DAGs.
Given a DAG $G= \tup{V,E}$ we ask if there is a 
partition of $V$ to $V_1$ and $V_2$ and $C = \{(u,v) \in E~ |~ u \in V_1, v \in V_2 \}$
so that $|C| \geq k$. 
From $G$, we create a TDAG $G'$ by adding the source vertex 
$s$ and the sink vertex $t$, which we connect with all of $V$. 
$G$ has a directed cut $C$ of size $k$ iff $G'$ has an $s-t$ directed 
cut $S$ of size $|V|+k$. 
Let $S$ be the union of $C$ with the edges from $V_1$ to $t$ and the 
edges from $s$ to  $V_2$. 
By construction, $S$ is a minimal $s-t$ cut and $|S|= |C| + |V|$.
%For the other direction, 
On the other hand, a minimal $s-t$ cut $S$ of size $|V|+k$ in $G'$ partitions the vertices 
depending on whether they are reachable from $s$ when $S$ is removed. Let $A$ be this set of
vertices and let $B$ contain the rest of them, excluding $t$. $S$ is a cut, so it must contain every edge from $A$ to $B$, every edge from $s$ to $B$, and every edge from $A$ to $t$.  This partition 
induces a cut in $G$ of size $|S|-(|A|+|B|)=k$.
\end{proof}}

\begin{proof}
\maxp problem is in \NP. Given any 2-terminal directed graph $G = \tup{V,E}$ and a set $S$ of 
edges in $E$ we can easily check whether $S$ is an $s-t$ cut; if $S$ is indeed an $s-t$ cut,
then by deleting the edges in $S$ there is no directed path from $s$ to $t$ and this can be 
easily verified via Dijkstra algorithm .

To show completeness we reduce from \maxdcut on DAGs~\cite{LKM11}. 
In an instance of \maxdcut  problem we are given a directed acyclic graph
$G= \tup{V,E}$  and an integer $k$, and we are asked if there a partition of $V$ into two sets
$V_1$ and $V_2$ so that the cardinality of the edge set 
$C = \{(u,v) \in E | u \in V_1, v \in V_2 \}$ is at least $k$. We construct a 2-terminal DAG
$G'$ as %we did in the proof of Theorem~\ref{thm:lpath-general}. 
follows.
We add the vertex $s$ and we connect it with every vertex $v \in V$ via an edge directed from $s$ to $v$. 
Furthermore, we add the vertex $t$ and we connect it with every vertex $v\in V$ via an
edge directed from $v$ to $t$. Clearly, $G'$ is a 2-terminal graph. Furthermore, it is not
hard to see that no directed cycles were created. Thus, $G'$ is a 2-terminal DAG.
We will prove that there exists a directed cut of size $k$ in $G$ if and only if
there exists an $s-t$ directed cut  of size $|V|+k$ in $G'$. 

Firstly, assume that in $G$ there 
exists a partition of $V$ to $V_1$ and $V_2$ such that the size of $C$, i.e., the number of
directed edges from $V_1$ to $V_2$, is $k$. Then, the set $S$ that contains $C$, the 
edges from the vertices of $V_1$ to $t$ and the edges from $s$ to vertices of $V_2$, is
a minimal $s-t$ cut. Observe, $|S|=|C|+|V_1|+|V_2| = k + |V|$. To see why $S$ is an $s-t$
cut, observe that there is no path of the form $s-v-t$ with $v \in V$, because one of the 
edges $(s,v)$ and $(v,t)$ is missing. The only other way to reach $t$ from $s$ is to go 
from $s$ to some vertex of $V_1$, move to $V_2$, and then reach $t$. But every edge 
from $V_1$ to $V_2$ is  in $C$, hence there is no such $s-t$ path. Furthermore, $S$ is
minimal since for any edge $(u,v)$ in $C$ there is clearly a path $s-u-v-t$ in $G'$ that does not
contain any other edge in $S$, and for any other edge in $S \setminus C$ there is an $s-t$
path of length three that does not use any other in $S$. 

For the other direction now, consider a minimal $s-t$ cut $S$ in $G'$ of size $|V|+k$. 
Denote by $A$ all the vertices accessible from $s$ in $E \setminus S$, and by $B$ all other
vertices of $G$. The cut $S$ contains every edge from $A$ to $B$, every edge from $s$
to $B$, and every edge from $A$ to $t$, so in particular we get that the size of the cut
defined by the partition of $V$ to $A$ and $B$ in $G$ is exactly 
$|S|-(|A|+|B|) = |V|+k-|V|=k$. Finally, observe that the partition defined by $A$ and $B$ is a
directed cut for $G$, because otherwise there would be a directed $s-t$ path and thus $S$
would not be an $s-t$ cut.
\end{proof}

As an immediate corollary we get that \isDMinor and \isDEmbedded are  \NP-complete even on a TDAG. When $G'=\tup{V',E'}$ is fixed, both problems  are in \PP: we use the 
algorithm of~\cite{ForHopWyl80} for h-embedding as a subroutine on at most $2^{|V'|^3}$ 
graphs due to Proposition~\ref{prop:embed_finite}.

\begin{proposition}\label{prop:isminor_TDAG_hard}
\isDMinor and \isDEmbedded are  \NP-complete even on a TDAG. 
\end{proposition}

\begin{proof}
By Prop.~\ref{th:PW_embed} checking the parallel width  (\maxp) of a TDAG is equivalent to check if it contains the d-minor $G_{P(k)}$. Thus, the proof for \isDMinor follows from Theorem~\ref{prop:maxp_TDAG_hard}. By Thm.~\ref{thm:minor_embed} \isDMinor and \isDEmbedded are equivalent on a TDAG.
\end{proof}

	\begin{theorem}\label{thm:isembedded} \isDEmbedded and \isDMinor are in \PP for TDAGs when $G'$ is fixed. 
	\end{theorem}
	\begin{proof}Denote $k=|V'|$. 
	We would like to use the algorithm of \cite{ForHopWyl80} (see Theorem~3 there) that checks if $G'$ is homeomorphic to a subgraph of $G$, i.e., if $G'$ is h-embedded in $G$. Since h-embedding does not allow vertex split operations, it is possible that $G'$ is d-embedded but not h-embedded in $G$. However, by Prop.~\ref{prop:embed_finite} there are at most $2^{k^2}$ graphs such that one of them is h-embedded in $G$, so we can try all of them. 
	%
	%
	%This can only happen if there are ``junction'' vertices in $G'$ that have both an indegree and an outdegree larger than one. We thus create a list of all TDAGs that can be obtained from $G'$ by splitting a subset of junction vertices. Note that there are at most $2^k$ such graphs. If one of these graphs $G_j$ is h-embedded in $G$, then $G'$ is d-embedded in $G$. In the other direction, if there is a sequence of d-embedding operations from $G'$ to $G$, w.l.o.g. it starts with all split operations by Lemma~\ref{lemma:embed_subgraph}.\adl{fix?} Split operations on vertices with indegree or outdegree one are equivalent to edge subdivision, and can be performed after all split operations on junctions. However, after all split operations on junctions we get one of the $G_j$ graphs that is h-embedded in $G$. \adl{cannot understand the last sentence.}
	%
	%
Overall, we run the h-embedding algorithm of \cite{ForHopWyl80} (whose runtime is  $|V|^{O(|E'|)}= |V|^{O(k^2)}$) at most $2^{k^2}$ times, so the runtime of our algorithm for \isDEmbedded is still polynomial in $|V|$. 

 By Thm.~\ref{thm:minor_embed} this also settles \isDMinor.  
\end{proof}

\begin{proposition}\label{prop:isembedded_TDAG}
\isDEmbedded is in \PP for a fixed $G'$ that is a TDAG, even when $G$ is not a TDAG.
\end{proposition}

\begin{proof}
We first check if $G$ is acyclic. If not, then $G'$ cannot be d-embedded in $G$ by Lemma~\ref{lemma:embed_closed}. Otherwise, $G$ is a TDAG so we can use Theorem~\ref{thm:isembedded}.
\end{proof}

Since by Theorems~\ref{th:PW_embed} and \ref{th:SPW_embed} finding the parallel (or serial-parallel) width is equivalent to check for excluded minors whose size is a function of $k$, we get the following. 
\begin{corollary}\label{cor:minors}
\maxp and \maxsp are in \PP for TDAGs and fixed $k$.
\end{corollary}

%
%\begin{proposition}\label{prop:maxsp_np}
%\maxsp is in \NP.
%\end{proposition}
%\begin{proof}
%For containment, note that by Proposition~\ref{prop:PW_embed}, $\PW(G)\geq k$ if and only if $G$ contains $G_{P(k)}$ as a d-minor. Since \dminor is in \NP by Theorem~\ref{thm:minor-cycle-np}, so is \maxsp.
%\end{proof}
%\noijcai{
\begin{proposition}\label{prop:maxsc_np}
\maxsc is \NP-complete.
\end{proposition}
\begin{proof}
Containment follows since \maxs and \maxc are in \NP. 

For hardness, we use a reduction from \maxs. Given a 2-terminal graph $G=\tup{V,E,s,t}$, we add two new terminals $\{s',t'\}$, two edges $(s',x)$ and two edges $(x,t')$ for every node $x\in V$. We argue that every set of edges in $E$ is concurrent. Indeed, consider a set $S$ and an edge $e\in S$. The path composed of $e=(a,b)$ and the new edges $(s',a),(b,t')$ does not use any edge from $S\setminus \{e\}$. 

This means that for every path $p$ in $G$, the path $s'-[p]-t'$ in $G$ is a serial-concurrent set of size $|p|+2$. 

In the other direction, if there is a serial-concurrent set $S$ of size at least $k$ in $G'$, let $p$ be the $s'-t'$ path containing $S$. The path $p$ without vertices $s',t'$ is a simple path in $G$ of length as least $k-2$.   
\end{proof}
%}

%\input{minimal_cuts}
\if
\rmr{This is not true. The proof works only for serial and concurrent sets. The characterization of a \emph{parallel} set of edges $S=\{(a_i,b_i)\}_{i\leq k}$ is that there are vertex-disjoint components $T_s,T_t$, such that $T_s$ contains $s$ and all of $a_i$, and $T_t$ contains $t$ and all of $b_i$. I don't know how to check efficiently if such components exist.}
  
\begin{proposition}\label{prop:poly_parallel}Given a TDAG $G=\tup{V,E,s,t}$ of size $n$ and set of edges $S\subseteq E$ of size $k$, it can be checked in time polynomial in $n$ and $k$ whether $S$ is serial and parallel. 
\end{proposition}
\begin{proof}
Since $G$ is a TDAG, we can sort all vertices in topological sort. Sort all edges in $S$ according to their first vertex $a_1,a_2,\ldots,a_k$. Clearly if $e_1=(a_1,b_1)$, $e_2=(a_2,b_2)$, and $a_1$ precedes $a_2$, then edge $e_2$ cannot precede $e_2$ on any path. To check whether $S$ is serial, it remains to check if there is a path (including a 0-length path) from $s$ to $a_1$, from $b_1$ to $a_2$, and so on. 

To check whether $S$ is \rmr{concurrent}, we consider each edge $e_i\in S$, and check if there is an $s-t$ path that contains $e_i$ but not any other edge from $S$. To that end, we check if the graph $G'\tup{V,E\setminus S,s,t}$ contains the paths $s-a_i$ and $b_i-t$. 
\end{proof}
\begin{corollary}
Given a TDAG $G=\tup{V,E,s,t}$ of size $n$, it can be checked whether $SPW(G)<k$ in time polynomial in $n^k$. 
\end{corollary}
This is by applying  Prop.~\ref{prop:poly_parallel} to any subset of edges of size $<k$. In fact this is true even for cyclic graphs, since we can try all $k!$ permutations of $S$ to see if it is serial. 

Is there an algorithm that is polynomial in $k$ for general 2-terminal graphs? What about TDAGs? Note that to show $SPW(G)\geq k$ we should find $k$ edges that are both serial and parallel. 
Finding a serial set of size $k$ is essentially the Longest Path problem, which is NP-hard in general 2-terminal graphs but easy in TDAGs. Perhaps this can be used to show that checking $SPW(G)\geq k$ in general 2-terminal graphs is hard too.

 Finding a $k$-size parallel set (the Largest Minimal Cut, or Largest Bond) sounds like a natural problem but I did not find it in the literature. I suspect that like Longest path it is hard for general 2-terminal graphs and easy for TDAGs. In particular, I started to think of a dynamic algorithm that first sorts the vertices using topological sort and then tracks for every vertex $a_i,~~ i=1,2,3,\ldots$ the largest minimal $s-t$ cut the only contains edges whose vertices  are in $\{a_1,\ldots,a_i\}$.  
\fi
\section{Discussion}
Many different variations of operations can be used to obtain ``simple'' graphs that capture the essential forbidden properties of large classes of graphs: minors, embeddings, subdivisions, etc. These operations should be rich enough to allow for a small set of forbidden graphs, but restricted enough to only capture the intended class. %So there is some trade off between the efficiency 

\noijcai{
For example, s-embedding, which is richer than other embeddings for undirected graphs, allows for a characterization of extension-parallel networks with just 1 forbidden graph instead of 3 (see \cite{milchtaich2006network,epstein2009efficient}). One can check that applying s-embeddings to other characterizations can reduce the number of required forbidden graphs from 5 (in Lemma~2 of \cite{milchtaich2015network}) to 2, or from 9 (in Proposition~2 of \cite{acemoglu2016informational}) to 2. Also, for each of the graphs $K_{5}$ and $K_{3,3}$ used in the celebrated characterization of planar graphs~\cite{wagner1937eigenschaft},  9 more graphs would be required if we only allow edge subdivision rather than node splits (the reverse operation of edge contraction). 
}

We believe that d-embeddings and d-minors 
%(which coincide for directed 2-terminal acyclic graphs)
 will turn out to be useful, beyond the applications demonstrated in the paper. 
%An indication for this is that they simplify known characterizations. For example, multiextension-parallel graphs characterizations with subdivision graphs require two forbidden graphs~\cite{holzman2015strong}, but one of them is d-embedded in another other, and it is easy to verify that these two graphs suffice to characterize the class with d-embeddings. To characterize acyclic graphs, only one graph is required (a bidirectional edge).
For example, in \cite{kleinberg2014time} bad graphs for planning are identified by \emph{undirected minors}, which mislabels many graphs due to ignoring edge directions. A tighter characterization could be obtained by d-minors.  

It is interesting whether d-embeddings or d-minors can be used to characterize other classes of directed graphs, such as graphs with bounded triangular width~\cite{meer2011extended} or D-width~\cite{safari2005d}. %, or directed path-width~\cite{barat2006directed}. %, or DAG-width~\cite{obdrvzalek2006dag}. 
Finally, there is the question of whether a directed graph version of the Graph Minor Theorem holds for d-minors or 
d-embeddings~\cite{kawarabayashi2015towards}.

%An open challenge is to find a richer notion than d-embedding, that will enable us to characterize graphs with parallel-width of at most $k$ with a polynomial (or even constant) number of forbidden graphs. 

 %
%
%
%For undirected graphs, \cite{acemoglu2016informational} characterize series-linearly independent graphs (extension-parallel graphs connected in sequence) by a set of 9 graphs that must not be embedded (Proposition~2 there). However, one of these graphs is \emph{s-embedded} in 7 of the others.  
%
%Similarly, in Lemma~2 of \cite{milchtaich2015network}, networks with a property called \emph{topological existence} are characterized by a set of 5 graphs that must not be embedded (with a slightly different definition of embedding). Once again, 2 of this graphs are s-embedded in the other three.  

\section*{Acknowledgments}
%\begin{acks}
The authors thank Ron Holzman, Pieter Kleer, Igal Milchtaich,  Evdokia Nikolova, David Parkes, Yuval Salant, Nicol\'as Stier-Moses and \'Eva Tardos for  helpful discussions and comments.

\begin{small}
\bibliographystyle{plainurl}
\bibliography{smoothness-short}
\end{small}

\end{document}

%% file: defs.tex
\newcommand{\coursename}{(67686) Mathematical Foundations of AI}

\newcommand{\handout}[5]{
   \renewcommand{\thepage}{#1-\arabic{page}}
   \noindent
   \begin{center}
   \framebox{
      \vbox{
    \hbox to 5.78in { {\bf \coursename}
         \hfill #2 }
       \vspace{4mm}
       \hbox to 5.78in { {\Large \hfill #5  \hfill} }
       \vspace{2mm}
       \hbox to 5.78in { {\it #3 \hfill #4} }
      }
   }
   \end{center}
   \vspace*{4mm}
}

\def\({\left(}
\def\){\right)}

\def\ol{\overline}

\newenvironment{proof-sketch}{\noindent{\bf Sketch of Proof}\hspace*{1em}}{\qed\bigskip}
\newenvironment{proof-idea}{\noindent{\bf Proof Idea}\hspace*{1em}}{\qed\bigskip}
\newenvironment{proof-of-lemma}[1]{\noindent{\bf Proof of Lemma #1}\hspace*{1em}}{\qed\bigskip}
\newenvironment{proof-attempt}{\noindent{\bf Proof Attempt}\hspace*{1em}}{\qed\bigskip}

\newcommand{\FOR}{{\bf for}}

\newcommand{\IF}{{\bf if}}

\makeatletter
\def\fnum@figure{{\bf Figure \thefigure}}
\def\fnum@table{{\bf Table \thetable}}
\long\def\@mycaption#1[#2]#3{\addcontentsline{\csname
  ext@#1\endcsname}{#1}{\protect\numberline{\csname
  the#1\endcsname}{\ignorespaces #2}}\par
  \begingroup
    \@parboxrestore
    \small
    \@makecaption{\csname fnum@#1\endcsname}{\ignorespaces #3}\par
  \endgroup}
\def\mycaption{\refstepcounter\@captype \@dblarg{\@mycaption\@captype}}
\makeatother

\newcommand{\mathify}[1]{\ifmmode{#1}\else\mbox{$#1$}\fi}
\newcommand{\bigO}O

\newcommand\tup[1]{\left\langle #1 \right\rangle}

\def\PW{\mbox{PW}}

\newcommand{\remove}[1]{{}}

%% file: no_directed_embed.tex
%\documentclass[]{article}
%\usepackage{tikz}
%\usepackage{tikz}%,fullpage}
%\usetikzlibrary{arrows,%
                %petri,%
                %topaths}%
%\usepackage{tkz-graph}
%\begin{document}
%\begin{figure}
%%%%%%%%%%%%%%%%%%%%%%
 \begin{minipage}[b]{0.4\textwidth}
\tikzstyle{VertexStyle}=[circle,draw=black,fill=white,
inner sep=0pt,minimum size=8mm]
\tikzset{
  font={\fontsize{18pt}{12}\selectfont}}
  \begin{center}
\begin{tikzpicture}[scale=0.6,transform shape]

  \Vertex[L=${s}$,x=0,y=6]{s}
	\Vertex[L=${a}$,x=-2,y=3]{a}
  \Vertex[L=${t}$,x=0,y=0]{t}
	\Vertex[L=${b}$,x=2,y=3]{b}
	\Vertex[L=${x}$,x=0,y=4]{x}
	\Vertex[L=${y}$,x=0,y=2]{y}

	  \tikzstyle{VertexStyle}=[rotate=-90,above]
	\tikzstyle{VertexStyle}=[fill=black!20!white]
\tikzstyle{VertexStyle}=[shape=coordinate]
%	\Vertex[x=-2,y=3]{e0}
  \tikzstyle{LabelStyle}=[fill=white,sloped]
  \tikzstyle{EdgeStyle}=[->,thin]

	\Edge(s)(a)
	\Edge(s)(b)
  \Edge(a)(x)
	\Edge(x)(b)
	\Edge(b)(y)
	\Edge(y)(a)
	\Edge(a)(t)
	\Edge(b)(t)
 
	%\tikzstyle{EdgeStyle}=[bend left]
	  %\Edge(t2)(s2)
		 %

  \tikzstyle{EdgeStyle}=[dashed]
	\Edge(x)(y)
%  \Edge[label=$630$]{S}{B}
%  \Edge[label=$210$]{S}{N}
%  \Edge[label=$230$]{S}{M}
\end{tikzpicture}
  \end{center}
\subcaption{}
\label{sfig:no_directed_embed}
\end{minipage}
%
%\end{figure}
%\end{document}

%% file: no_subdivide.tex
%\documentclass[]{article}
%\usepackage{tikz}
%%\usepackage{tikz}%,fullpage}
%\usetikzlibrary{arrows,%
                %petri,%
                %topaths}%
%\usepackage{tkz-berge}
%\usepackage[position=top]{subfig}
%\begin{document}
%\begin{figure}
\tikzstyle{VertexStyle}=[circle,draw=black,fill=white,
inner sep=0pt,minimum size=8mm]
\centering
\tikzset{
  font={\fontsize{18pt}{12}\selectfont}}
\begin{minipage}[b]{0.4\textwidth}
\label{sfig:subdivide}
\begin{tikzpicture}[scale=0.6,transform shape]
 %\tikzstyle{VertexStyle}=[]
  \Vertex[x=2,y=6,L=$s$]{s}
	\Vertex[x=2,y=3,L=$u$]{x}
  \Vertex[x=2,y=0,L=$t$]{t}

	%\AssignVertexLabel{s}{$s'$,$s''$}
  
  \tikzstyle{LabelStyle}=[fill=white,sloped]
  \tikzstyle{EdgeStyle}=[->,thin,bend right]
  %\Edge[label=$e_1$](s)(x)
	%\Edge[label=$e_2$](x)(t)
\Edge(s)(x)
	\Edge(x)(t)

\tikzstyle{EdgeStyle}=[->,thin,bend left]
  \Edge(s)(x)
	\Edge(x)(t)
	
\end{tikzpicture}
%}
~~~
%\subfloat[The graph $G'$]{ \label{pq_right}
\begin{tikzpicture}[scale=0.6,transform shape]
%\tikzstyle{VertexStyle}=[]
   \Vertex[x=2,y=6,L=$s$]{s}
	\Vertex[x=2,y=4,L=$u$]{x}
	\Vertex[x=2,y=2,L=$v$]{y}
  \Vertex[x=2,y=0,L=$t$]{t}

	%\AssignVertexLabel{s}{$s'$,$s''$}

  \tikzstyle{EdgeStyle}=[->,double]
\Edge(x)(y)
  
  \tikzstyle{LabelStyle}=[fill=white,sloped]
  \tikzstyle{EdgeStyle}=[->,thin,bend right]
  \Edge(s)(x)
	\Edge(y)(t)
\tikzstyle{EdgeStyle}=[->,thin,bend left]
  \Edge(s)(x)
	\Edge(y)(t)
\end{tikzpicture}
\subcaption{}
\label{sfig:subdivide}
\end{minipage}
~~~~~~~~~~~~~~~
\begin{minipage}[b]{0.4\textwidth}
\begin{tikzpicture}[scale=0.6,transform shape]
 %\tikzstyle{VertexStyle}=[]
  \Vertex[x=2,y=6,L=$s$]{s}
	\Vertex[x=0,y=3,L=$a$]{a}
  \Vertex[x=4,y=3,L=$b$]{b}
	\Vertex[x=2,y=0,L=$t$]{t}

	%\AssignVertexLabel{s}{$s'$,$s''$}
  
  \tikzstyle{LabelStyle}=[fill=white,sloped]
  \tikzstyle{EdgeStyle}=[->,thin]
  \Edge(s)(a)
	\Edge(s)(b)
	\Edge(a)(b)
	\Edge(a)(t)
	\Edge(b)(t)
	
\end{tikzpicture}
\subcaption{The Braess graph}
\label{sfig:Braess}
\end{minipage}
%\end{figure}
%
%\end{document}

%% file: network_fig_SPk_variants.tex
%\documentclass[]{article}
%\usepackage{tikz}
%\usepackage{tikz}%,fullpage}
%\usetikzlibrary{arrows,%
                %petri,%
                %topaths}%
%\usepackage{tkz-graph}
%\begin{document}
%\begin{figure}
%%%%%%%%%%%%%%%%%%%%%%
\centering
\tikzset{
  font={\fontsize{18pt}{12}\selectfont}}
\begin{tikzpicture}[scale=0.6,transform shape]

  \Vertex[L=$b_1$,x=0,y=1.5]{b1}
  %\Vertex[x=0,y=3]{a2}
  \Vertex[L=$a_2$,x=2,y=4.5]{a2}
  \Vertex[L=$b_2$,x=2,y=1.5]{b2}
	\Vertex[L=$a_3$,x=4,y=4.5]{a3}
	\Vertex[L=$b_3$,x=4,y=1.5]{b3}
  \Vertex[L=$a_{4}$,x=6,y=4.5]{a4}
	\Vertex[L=$b_{4}$,x=6,y=1.5]{b4}
	\Vertex[L=$a_{5}$,x=8,y=4.5]{a5}
	%\Vertex[L=$b_{5}$,x=8,y=1.5]{b5}
	%\Vertex[L=$a_{6}$,x=10,y=4.5]{a6}
	\Vertex[L=${s=a_1}$,x=0,y=6.5]{s}
  \Vertex[L=${t=b_5}$,x=8,y=-0.5]{t}

  \tikzstyle{VertexStyle}=[fill=black!20!white]
\tikzstyle{VertexStyle}=[shape=coordinate]
%	\Vertex[x=-2,y=3]{e0}
  \tikzstyle{LabelStyle}=[fill=white,sloped]

 \tikzstyle{EdgeStyle}=[->,thin]
	\Edge(s)(a2)
	\Edge(s)(a3)
  \Edge(s)(a4)
 \Edge(s)(a5)
 %\Edge(s)(a6)

\Edge(b1)(t)
\Edge(b2)(t)
	\Edge(b3)(t)
	\Edge(b4)(t)
	%\Edge(b5)(t)
	
	  \tikzstyle{EdgeStyle}=[->,double]
	\Edge(s)(b1)
	\Edge(a2)(b2)
	\Edge(a3)(b3)
	\Edge(a4)(b4)
 \Edge(a5)(t)
	%\Edge(a6)(t)

	\Edge(b1)(a2)
	\Edge(b2)(a3)
	\Edge(b3)(a4)
\Edge(b4)(a5)

	\tikzstyle{EdgeStyle}=[bend right]
	%			\Edge[label=$c_{st}$](s)(e0)
	\tikzstyle{EdgeStyle}=[->,bend right]

%	\Edge(e0)(t)
	
	\tikzstyle{EdgeStyle}=[->,bend left]

%  \tikzstyle{EdgeStyle}=[bend right]
%  \Edge[label=$630$]{S}{B}
%  \Edge[label=$210$]{S}{N}
%  \Edge[label=$230$]{S}{M}
\end{tikzpicture}
~~~~~
\begin{tikzpicture}[scale=0.6,transform shape]

  \Vertex[L=$b_1$,x=0,y=1.3]{b1}
  %\Vertex[x=0,y=3]{a2}
  \Vertex[L=$a_2$,x=2,y=4.5]{a2}
  \Vertex[L=$b_2$,x=2,y=1.5]{b2}
	\Vertex[L=$a_3$,x=4,y=4.5]{a3}
	\Vertex[L=$b_3$,x=4,y=1.6]{b3}
  \Vertex[L=$a_{4}$,x=6,y=4.5]{a4}
	\Vertex[L=$b_{4}$,x=6,y=0.5]{b4}
	\Vertex[L=$a_{5}$,x=8,y=4.5]{a5}
	%\Vertex[L=$b_{5}$,x=8,y=1.5]{b5}
	%\Vertex[L=$a_{6}$,x=10,y=4.5]{a6}
	\Vertex[L=${s=a_1}$,x=0,y=6.5]{s}
  \Vertex[L=${t=b_5}$,x=8,y=-0.5]{t}

  \tikzstyle{VertexStyle}=[fill=black!20!white]
\tikzstyle{VertexStyle}=[shape=coordinate]
%	\Vertex[x=-2,y=3]{e0}
  \tikzstyle{LabelStyle}=[fill=white,sloped]

 \tikzstyle{EdgeStyle}=[->,thin]
	\Edge(s)(a2)
	\Edge(a2)(a3)
  \Edge(s)(a4)
 \Edge(a4)(a5)
 %\Edge(s)(a6)

\Edge(b1)(t)
\Edge(b2)(b4)
	\Edge(b3)(b4)
	\Edge(b4)(t)
	%\Edge(b5)(t)
	
	  \tikzstyle{EdgeStyle}=[->,double]
	\Edge(s)(b1)
	\Edge(a2)(b2)
	\Edge(a3)(b3)
	\Edge(a4)(b4)
 \Edge(a5)(t)
	%\Edge(a6)(t)

	\Edge(b1)(a2)
	\Edge(b2)(a3)
	\Edge(b3)(a4)
\Edge(b4)(a5)

	\tikzstyle{EdgeStyle}=[bend right]
	%			\Edge[label=$c_{st}$](s)(e0)
	\tikzstyle{EdgeStyle}=[->,bend right]

%	\Edge(e0)(t)
	
	\tikzstyle{EdgeStyle}=[->,bend left]

%  \tikzstyle{EdgeStyle}=[bend right]
%  \Edge[label=$630$]{S}{B}
%  \Edge[label=$210$]{S}{N}
%  \Edge[label=$230$]{S}{M}
\end{tikzpicture}
%\end{figure}
%\end{document}

%% file: no_d_embed.tex
%\documentclass[]{article}
%\usepackage{tikz}
%%\usepackage{tikz}%,fullpage}
%\usetikzlibrary{arrows,%
                %petri,%
                %topaths}%
%\usepackage{tkz-berge}
%\usepackage[position=top]{subfig}
%\begin{document}
%\begin{figure}
\centering
\subfloat[$G'$ and $G$]{ \label{d_embed}
\begin{tikzpicture}[scale=0.6,transform shape]
 %\tikzstyle{VertexStyle}=[]
  \Vertex[x=0,y=3]{a}
  \Vertex[x=4,y=3]{b}
	\Vertex[x=2,y=6]{s}
	\Vertex[x=2,y=3]{x}
  \Vertex[x=2,y=0]{t}

	%\AssignVertexLabel{s}{$s'$,$s''$}
  
  \tikzstyle{LabelStyle}=[fill=white,sloped]
  \tikzstyle{EdgeStyle}=[->,thin]
  \Edge(s)(a)
	\Edge(s)(x)
	\Edge(s)(b)
  \Edge(a)(x)
	\Edge(x)(b)
	\Edge(a)(t)
	\Edge(x)(t)
	\Edge(b)(t)
\end{tikzpicture}
%}

%\subfloat[The graph $G'$]{ \label{pq_right}
\begin{tikzpicture}[scale=0.6,transform shape]
%\tikzstyle{VertexStyle}=[]
    \Vertex[x=0,y=3]{a}
  \Vertex[x=4,y=3]{b}
	\Vertex[x=2,y=6]{s}
	\Vertex[x=2,y=4]{x}
	\Vertex[x=2,y=2]{y}
  \Vertex[x=2,y=0]{t}

	%\AssignVertexLabel{s}{$s'$,$s''$}
  
  \tikzstyle{LabelStyle}=[fill=white,sloped]
  \tikzstyle{EdgeStyle}=[->,thin]
  \Edge(s)(a)
	\Edge(s)(x)
	\Edge(s)(b)
  \Edge(a)(y)
	\Edge(x)(b)
	\Edge(a)(t)
	\Edge(y)(t)
	\Edge(b)(t)
	 \tikzstyle{EdgeStyle}=[->,double]
	\Edge(x)(y)
\end{tikzpicture}
}
~~~~~~~~
\subfloat[$\overline G'$ and $\overline G$]{ \label{s_embed}
\begin{tikzpicture}[scale=0.6,transform shape]
 %\tikzstyle{VertexStyle}=[]
  \Vertex[x=0,y=3]{a}
  \Vertex[x=4,y=3]{b}
	\Vertex[x=2,y=6]{s}
	\Vertex[x=2,y=3]{x}
  \Vertex[x=2,y=0]{t}

	%\AssignVertexLabel{s}{$s'$,$s''$}
  
  \tikzstyle{LabelStyle}=[fill=white,sloped]
  \tikzstyle{EdgeStyle}=[thin]
  \Edge(s)(a)
	\Edge(s)(x)
	\Edge(s)(b)
  \Edge(a)(x)
	\Edge(x)(b)
	\Edge(a)(t)
	\Edge(x)(t)
	\Edge(b)(t)
\end{tikzpicture}
%}

%\subfloat[The graph $G'$]{ \label{pq_right}
\begin{tikzpicture}[scale=0.6,transform shape]
%\tikzstyle{VertexStyle}=[]
    \Vertex[x=0,y=3]{a}
  \Vertex[x=4,y=3]{b}
	\Vertex[x=2,y=6]{s}
	\Vertex[x=2,y=4]{x}
	\Vertex[x=2,y=2]{y}
  \Vertex[x=2,y=0]{t}

	%\AssignVertexLabel{s}{$s'$,$s''$}
  
  \tikzstyle{LabelStyle}=[fill=white,sloped]
  \tikzstyle{EdgeStyle}=[thin]
  \Edge(s)(a)
	\Edge(s)(x)
	\Edge(s)(b)
  \Edge(a)(y)
	\Edge(x)(b)
	\Edge(a)(t)
	\Edge(y)(t)
	\Edge(b)(t)
	 \tikzstyle{EdgeStyle}=[double]
	\Edge(x)(y)
\end{tikzpicture}
}
%\end{figure}

%\end{document}